\documentclass[pre,twocolumn,twoside,byrevtex,superscriptaddress,floatfix]{revtex4-1}

\usepackage [english]{babel}
\usepackage{indentfirst}
\usepackage[section]{placeins}
\usepackage{amssymb}
\usepackage{gensymb}
\usepackage{amsmath}
\usepackage{graphics}
\usepackage{booktabs}
\usepackage{lipsum}
\usepackage{multirow}
  \newcommand{\tb}{\color{black}}

\usepackage{currfile}
\usepackage{graphicx,epsfig,verbatim,enumerate}
\usepackage{amssymb,amsmath}
\usepackage{ifthen}

\usepackage{longtable}

\usepackage{mathtools}

\newboolean{twocolswitch}

\newcommand{\sindex}[1]{}
\newcommand{\nindex}[1]{}

\newcommand{\www}[1]{\url{#1}}

\usepackage{lettrine}

\usepackage{color}

\setboolean{twocolswitch}{true}

\begin{document}

\title{\protect
Transitions in climate and energy discourse between Hurricanes Katrina and Sandy
}

\author{
\firstname{Emily M.}
\surname{Cody}
}
\email{emily.cody@uvm.edu}
\affiliation{Department of Mathematics \& Statistics,
  Vermont Complex Systems Center,
  Computational Story Lab,
  \& the Vermont Advanced Computing Core,
  The University of Vermont,
  Burlington, VT 05405.}

\author{
\firstname{Jennie C.}
\surname{Stephens}
}
\email{jennie.stephens@uvm.edu}
  \affiliation{Rubenstein School of Environment and Natural Resources, University of Vermont, Burlington VT 05401}

\author{
\firstname{James P.}
\surname{Bagrow}
}
\email{james.bagrow@uvm.edu}
\affiliation{Department of Mathematics \& Statistics,
  Vermont Complex Systems Center,
  Computational Story Lab,
  \& the Vermont Advanced Computing Core,
  The University of Vermont,
  Burlington, VT 05405.}
  
 \author{
\firstname{Peter Sheridan}
\surname{Dodds}
}
\email{peter.dodds@uvm.edu}
\affiliation{Department of Mathematics \& Statistics,
  Vermont Complex Systems Center,
  Computational Story Lab,
  \& the Vermont Advanced Computing Core,
  The University of Vermont,
  Burlington, VT 05405.}
  
    \author{
\firstname{Christopher M.}
\surname{Danforth}
}
\email{chris.danforth@uvm.edu}
\affiliation{Department of Mathematics \& Statistics,
  Vermont Complex Systems Center,
  Computational Story Lab,
  \& the Vermont Advanced Computing Core,
  The University of Vermont,
  Burlington, VT 05405.}

\date{\today}

\begin{abstract}
  \protect
  Although climate change and energy are intricately linked, their explicit connection is not always prominent in public discourse and the media. Disruptive extreme weather events, including hurricanes, focus public attention in new and different ways, offering a unique window of opportunity to analyze how a focusing event influences public discourse.  Media coverage of extreme weather events simultaneously shapes and reflects public discourse on climate issues.  Here we analyze climate and energy \tb{newspaper} coverage of Hurricanes Katrina (2005) and Sandy (2012) using topic models, mathematical techniques used to discover abstract topics within a set of documents.  Our results demonstrate that post-Katrina media coverage does not contain a climate change topic, and the energy topic is limited to discussion of energy prices, markets, and the economy with almost no explicit linkages made between energy and  climate change.  In contrast, post-Sandy media coverage does contain a prominent climate change topic, a distinct energy topic, as well as integrated representation of climate change and energy, indicating a shift in climate and energy reporting between Hurricane Katrina and Hurricane Sandy.
 
\end{abstract}

\pacs{89.65.-s,89.75.Da,89.75.Fb,89.75.-k}

\maketitle

\section{Introduction}

Climate change is one of the most challenging issues of our time.  Anticipated climate disruptions, including a 4\degree C increase in the Earth's average temperature by the end of the 21st century \cite{ipcc2014} and more frequent and intense extreme weather events, result from increased atmospheric concentrations of greenhouse gases attributed primarily to fossil fuel burning for energy.  

Given probable links between the increasing ocean temperature and the severity and frequency of hurricanes and tropical storms \cite{mann2006atlantic, field2012managing, huber2011extreme}, extreme weather events have potential to raise awareness and increase public concern about climate change.  The disruptions caused by hurricanes and other storms can also raise awareness and focus attention on energy system vulnerability.  These extreme events can serve as a teachable experience for those not previously engaged with these issues \cite{myers2013relationship}.  Indeed, previous research has shown that after experiencing a large hurricane, citizens are more likely to adopt a pro-environmental belief system and support politicians who are climate change activists \cite{rudman2013truth}.  Populations living as far as 800 km from the path of a hurricane report having experienced it in some way \cite{howe2014mapping}.  Extensive news coverage of extreme weather events has also been found to increase public awareness of climate change by highlighting tangible and specific risks \cite{bell1994media, wilson2000drought}.  It has also been shown that individuals affected by a natural disaster are more likely to strengthen interactions on social media \cite{phan2015natural}.  As climate change news is prominent on social media \cite{cody2015climate}, these interactions provide another mechanism for raising climate change awareness following a natural disaster.    

This research recognizes the complex relationship between the news media and public discourse on science and policy. The news media both shapes public perceptions and public discourse and reflects and represents public perceptions and public discourse \cite{graber2009mass, gamson1989media}.  The media shapes public opinion of science by avoiding complex scientific language and displaying information for the layperson \cite{murray2001ain, peterson2009environmental, priest2009doing}.  People are more likely to learn about environmental and other science related risks through the media than through any other source \cite{corbett2004testing, peterson2009environmental}.  Research indicates that news media establish the context within which future information will be interpreted \cite{peterson2009environmental}.  In this research we analyze media coverage to characterize differences in the public discourse about climate change and energy after Hurricane Katrina and Hurricane Sandy.

Links between climate change and energy are often focused on climate mitigation, e.g., reducing greenhouse gas emissions from energy systems by shifting low-carbon energy systems.  However, climate change and energy are also linked in terms of increased energy system vulnerability in a changing climate \cite{stephens2013getting}.  Hurricanes and other extreme weather events often cause disruptions to energy systems including infrastructure damage, fuel supply shortages, and increases in energy prices.  Flooding and high wind speeds reveal multiple energy system vulnerabilities including evacuations of oil rigs and power outages at refineries, which can contribute to energy supply shortages and price increases. 

Despite the multiple linkages between climate change and energy systems, the issues of climate and energy are still often discussed in the media separately \cite{stephens2009wind, wilson2009carbon}.  Greater integration of the public discourse on climate change and energy could facilitate more sophisticated consideration of the opportunities for changing energy systems to prepare for climate change \cite{ipcc2014,metz2009controlling}.

A 2005 study on climate change in the media revealed that articles often frame climate change as a debate, controversy, or uncertainty, which is inconsistent with how the phenomenon is framed within the scientific community \cite{antilla2005climate}.  A recent 2015 linguistic study determined that the IPCC summaries, intended for non-scientific audiences, are becoming increasingly more complex and more difficult for people to understand \cite{barkemeyer2015a}, which highlights the critical interpretive role of the media in public discourse.   

Here, we quantitatively compare media coverage of climate change, energy, and the links between climate and energy after Hurricanes Katrina and Sandy, two of the most disruptive and costly hurricanes to ever hit the United States \cite{knabb2006tropical,sandy}.  Since energy system disruption represents a tangible consequence of climate change, the linking of these two topics in post-hurricane newspaper coverage provides readers with a portal for climate change education and awareness.  Newspaper media was selected for analysis rather than social media because in the rapidly changing media landscape the circulation patterns of these well-established newspapers have been relatively stable during the study period.  Also, a 2014 study by the American Press Institute determined that 61\% of Americans follow the news through print newspapers and magazines alone.  69\% of Americans use laptops and computers which includes online newspapers.  88\% of Americans find their news directly from a news organization, as opposed to roughly 45\% from social media and 30\% from electronic news ads \cite{API}.  With this high percentage of Americans getting news from the media, analysis of climate change reporting provides insights on shifts in public discourse and awareness.

We apply two topic modeling techniques stemming from different areas of mathematics to a corpus (collection of text) of newspaper articles about each hurricane.  A topic model uses word frequencies within a corpus to assign one or more topics to each text. For our present analysis, we employ Latent Semantic Analysis (LSA), which uses singular value decomposition to reduce a term-document matrix to latent semantic space, and Latent Dirichlet Allocation (LDA), a probabilistic bayesian modeling technique, which defines each hidden topic as a probability distribution over all of the words in the corpus (we provide more details in the  Methods section, Sec.~\ref{methods}).  

We apply a topic modeling approach as a way to assess the integration of climate change, energy and the links between climate and energy within post-hurricane media coverage.  Topic modeling is a valuable tool for the kind of research we perform as it does not require manual coders to read thousands of articles.  Instead, a specified number of topics are determined through analysis of the frequency of each word in each article in the corpus.  The resulting model explains the corpus in detail by categorizing the articles and terms into topics.  

We focus on the two most disruptive and costly hurricanes in U.S. history.  In August 2005, Hurricane Katrina struck Louisiana as a Category 3 storm, affecting the Gulf Coast from central Florida to Texas, causing over 100 billion dollars in damage and roughly 1,800 deaths.  Katrina destroyed or severely damaged much of New Orleans and other heavily populated areas of the northern Gulf Coast, resulting in catastrophic infrastructure damage and thousands of job losses \cite{knabb2006tropical}.  Hurricane Sandy hit the northeastern United States in October 2012.  It was the largest hurricane of the 2012 Atlantic hurricane season, caused 233 reported deaths, and over 68 billion dollars in damage to residential and commercial facilities as well as transportation and other infrastructure \cite{sandy}.  Many businesses faced short term economic losses, while the travel and tourism industry experienced longer term economic difficulties.  In the time shortly after Sandy hit, repairs and reconstructions were estimated to take four years \cite{sandyecon}.

We use this quantitative approach to assess the degree to which climate change or energy related topics are included in newspaper coverage following Hurricanes Sandy and Katrina.  The individual words that define each topic reveal how climate change and energy were represented in post-event reporting, which in turn shapes public discourse. 

We first describe the dataset and methods of analysis in Sec.~\ref{methods}.  We then describe the results of each topic modeling technique for each hurricane and make comparisons between the two corpora in Sec.~\ref{results}. We explore the significance of these results in Sec.~\ref{discussion}\&\ref{conclusion}.    

\section{Methods}
\label{methods}

\subsection{Data Collection}

We collected newspaper articles published in major U.S. newspapers in the year following each of the hurricanes.  We chose the timespan of one year to capture the duration of media coverage following each hurricane and also to ensure we had enough articles from each hurricane to conduct a proper mathematical analysis. We identified newspaper articles through a search that included the name of the hurricane and either the word ``hurricane" or ``storm" in either the title or leading paragraphs of the article.  To account for regional variation in post-hurricane reporting, we chose four newspapers spanning major regions of the United States: Northeast, New England, Midwest, and West.  We chose the following four newspapers due to their high Sunday circulation, and because they are high-profile, established newspapers with high readership: The New York Times, The Boston Globe, The Los Angeles Times, and The Chicago Tribune are influential and well-respected nationally as well as locally. These four newspapers are consistently in the top 25 U.S. Sunday newspapers and were available for article collection through online databases.  We collected articles appearing onwards from the first of the month the hurricane occurred in throughout the subsequent year using the ProQuest, LexisNexis, and Westlaw Campus Research online databases.  The total number of articles collected and included in the corpora for analysis are 3,100 for Hurricane Katrina and 1,039 for Hurricane Sandy.  We transform each corpus into a term-document matrix for the analysis. 

\subsection{Latent Semantic Analysis}

\begin{figure}[h]
\centerline{\includegraphics[width=0.4\textwidth]{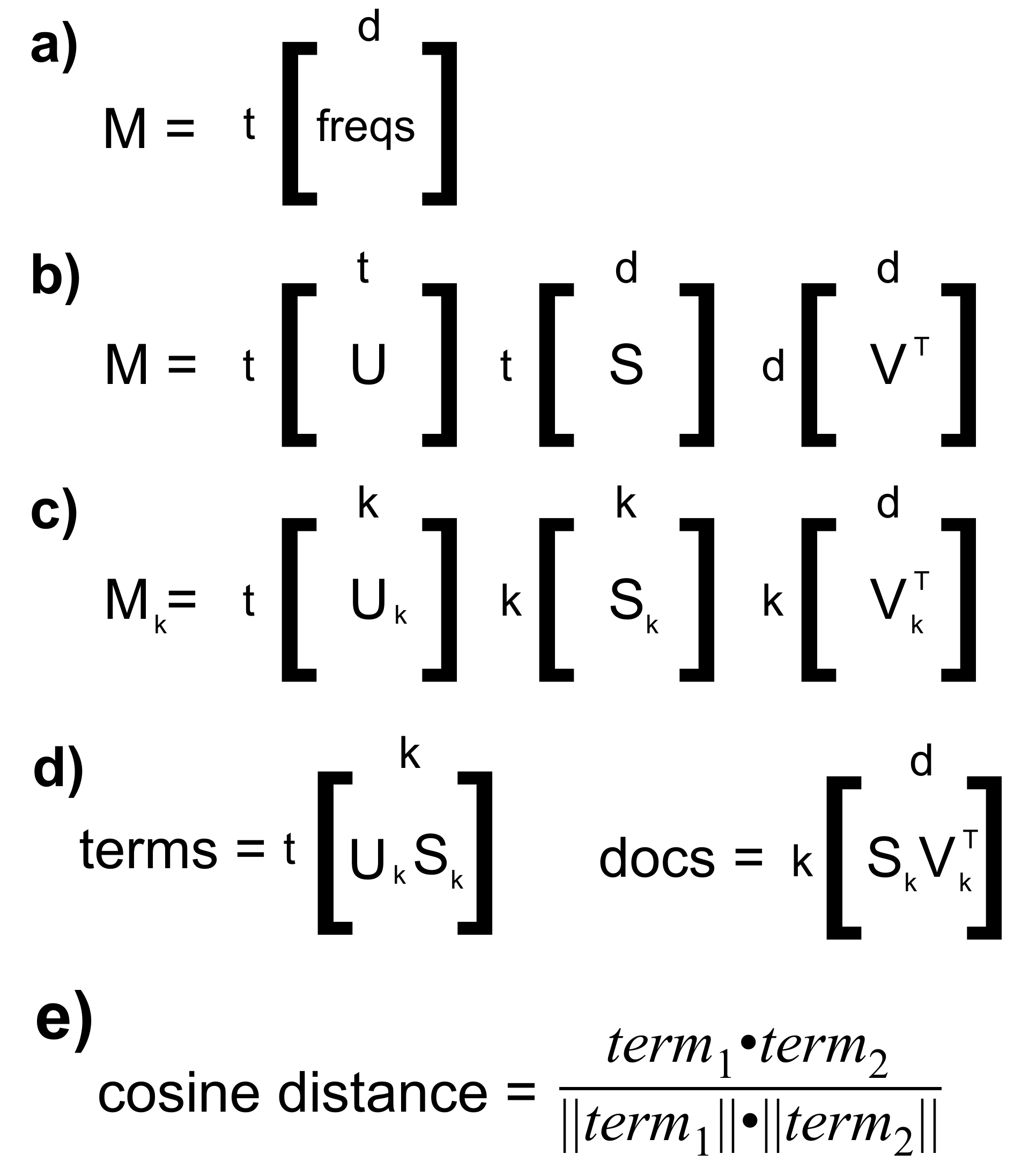}}
\caption{a) $M$ is a $t \times d$ matrix where $t$ and $d$ are the number of terms and documents in the corpus.  An entry in this matrix represents the number of times a specific term appears in a specific document. b) Singular Value Decomposition factors the matrix $M$ into three matrices.  The matrix $S$ has singular values on its diagonal and zeros everywhere else.  c) The best rank $k$ approximation of $M$ is calculated by retaining the $k$ highest singular values.  $k$ represents the number of topics in the corpus.  d) Each term and each document is represented as a vector in latent semantic space.  These vectors make up the rows of the term matrix and the columns of the document matrix.  e) Terms and documents are compared to each other using cosine similarity, which is determined by calculating the cosine of the angle between two vectors.}
\label{LSA}
\end{figure}

Latent Semantic Analysis (LSA) is a method of uncovering hidden relationships in document data \cite{deerwester1990indexing}.  LSA uses the matrix factorization technique Singular Value Decomposition (SVD) to reduce the rank of the term-document matrix, and merge the dimensions that share similar meanings.  SVD creates the following matrices:
$$M = USV^T,$$
where the matrix $M$ is the original $t\times d$ matrix \tb{(number of terms by number of documents)}, the columns of the matrix $U$ are the eigenvectors of $MM^T$, the entires in the diagonal of the matrix $S$ are the square roots of the eigenvalues of $MM^T$, and the rows of the matrix $V^T$ are the eigenvectors of $M^TM$.  Retaining the $k$ largest singular values and setting all others to 0 gives the best rank $k$ approximation of $M$.  This rank reduction creates a $\tb{t}\times k$ term matrix, $U_kS_k$, consisting of term vectors in latent semantic space as its columns, and a $k\times d$ document matrix, $S_kV_k^T$, consisting of document vectors as its rows.  The documents and terms are then compared in latent semantic space using cosine similarity as the distance metric \cite{berry2005understanding}.  If two term vectors have cosine distances close to 1, then these terms are interpreted to be related to each other in meaning.  We explain this process further in Fig.~\ref{LSA}.

We load the documents into a term-document matrix and remove common and irrelevant terms.  The terms we removed included terms common to the articles like ``hurricane", ``storm", ``sandy", and ``katrina", along with names of authors and editors of the articles.  We then convert each frequency in the matrix to term frequency-inverse document frequency (tf-idf) via the following transformation \cite{baeza1999modern}:
$$w_{i,j} = \left\{\begin{array}{ll}(1+\log_2 f_{i,j})\times\log_2\frac{N}{n_i} & f_{i,j} > 0 \\ 0 & \text{otherwise,}\end{array}\right.$$
where the variable $w_{i,j}$ is the new weight in the matrix at location $(i,j)$,  $f_{i,j}$ is the current frequency in position $(i,j)$,  $N$ is the number of documents in the corpus, and $n_i$ is the number of documents containing word $i$.  This weighting scheme places higher weights on rarer terms because they are more selective and provide more information about the corpus, while placing lower weights on common words such as ``the" and ``and".

We run LSA on the tf-idf term-document matrix for each hurricane.  We then compare the documents and terms in the corpus to a given query of terms in latent semantic space.  We transform the words that the query is composed of into term vectors, and calculate their centroid to give the vector representation of the query.  If the query is only one word in length, then the vector representation of the query equals the vector representation of the word.  We analyze three queries using LSA:  ``climate", ``energy", and ``climate, energy".  LSA gives the terms most related to this query vector, which we then use to determine how climate change and energy are discussed both separately and together in the media after Hurricanes Katrina and Sandy.

\subsection{Latent Dirichlet Allocation}

\begin{figure}[h]
\centerline{\includegraphics[width=0.5\textwidth]{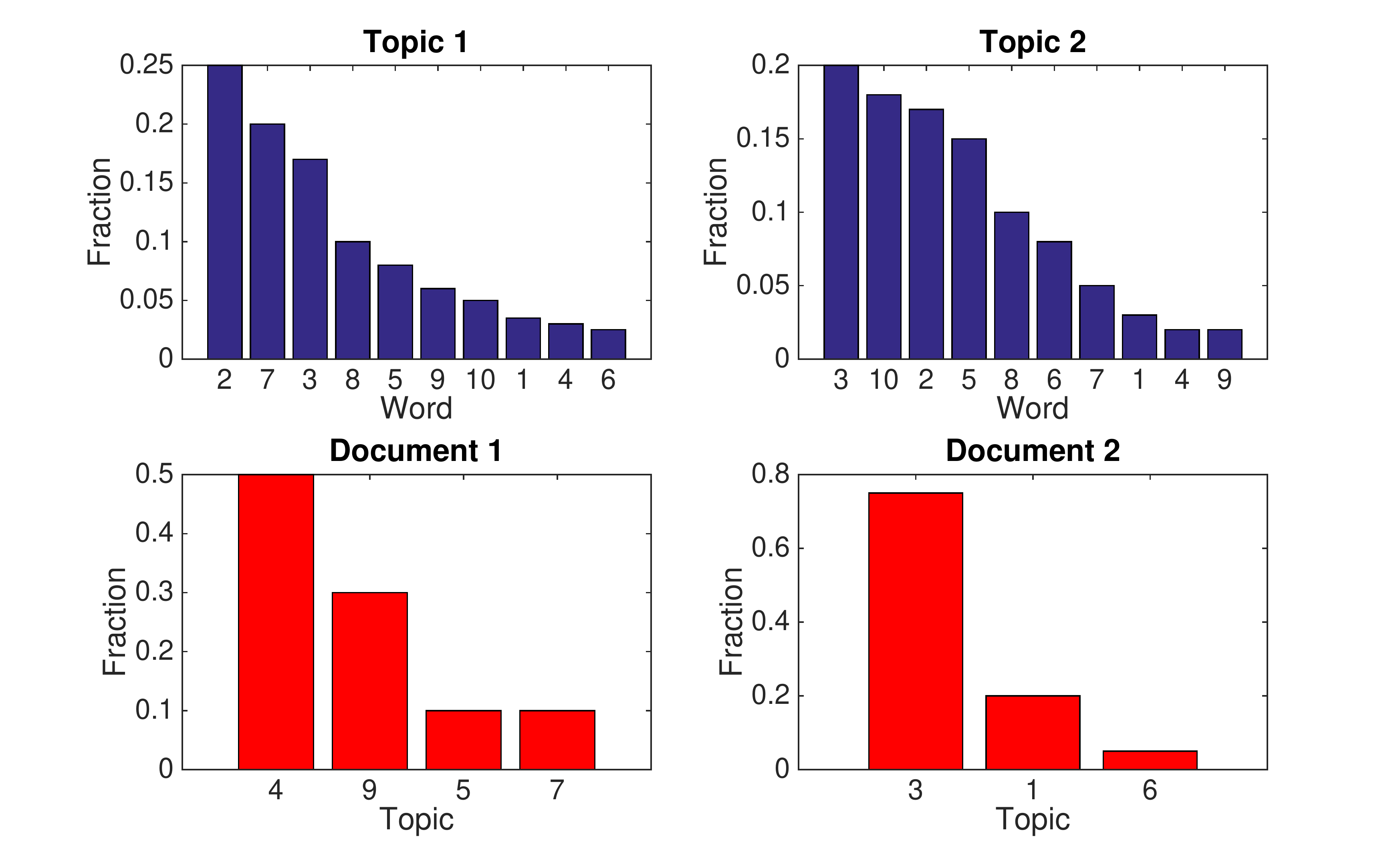}}
\caption{a) Examples of two topic distributions that may arise from an LDA model.  In this example, each topic is made up of 10 words and each word contributes to the meaning of the topic in a different proportion.  b) Examples of two document distributions that may arise from an LDA model.  Document 1 is made up of four major topics, while document 2 is made up of 3 major topics.}
\label{LDA}
\end{figure}

Latent Dirichlet Allocation (LDA), a probabilistic topic model \cite{blei2003latent, blei2012probabilistic}, defines each hidden topic as a probability distribution over all of the words in the corpus, and each document's content is then represented as a probability distribution over all of the topics.  Fig.~\ref{LDA} gives illustrations of distributions for a potential LDA model.

LDA assumes that the documents were created via the following generative process.  For each document: 
\begin{enumerate}
\item Randomly choose a distribution of topics from a dirichlet distribution.  This distribution of topics contains a nonzero probability of selecting each word in the corpus.
\item For each word in the current document:
\begin{enumerate}[a)]
\item Randomly select a topic from the topic distribution in part 1.
\item Randomly choose a word from the topic just selected and insert it into the document.
\end{enumerate}
\item Repeat until document is complete.
\end{enumerate} 
The distinguishing characteristic of LDA is that all of the documents in the corpus share the same set of $k$ topics, however each document contains each topic in a different proportion.  The goal of the model is to learn the topic distributions.  The generative process for LDA corresponds to the following joint distribution:
\begin{equation*}
\begin{aligned}
& P(\beta_{1:K},\theta_{1:D},z_{1:D},w_{1:D}) = \\
& \prod\limits_{i=1}^KP(\beta_i)\prod\limits_{d=1}^DP(\theta_d)\left(\prod\limits_{n=1}^NP(z_{d,n}|\theta_d)P(w_{d,n}|\beta_{1:K},z_{d,n})\right),
\end{aligned}
\end{equation*}
where $\beta_k$ is the distribution over the words, $\theta_{d,k}$ is the topic proportion for topic $k$ in document $d$, $z_{d,n}$ is the topic assignment for the $n$th word in document $d$, and $w_{d,n}$ is the $n$th word in document $d$.  This joint distribution defines certain dependences.  The topic selection, $z_{d,n}$ is dependent on the topic proportions each the article, $\theta_d$. The current word $w_{d,n}$ is dependent on both the topic selection, $z_{d,n}$ and topic distribution $\beta_{1:k}$. The main computational problem is computing the posterior.  The posterior is the conditional distribution of the topic structure given the observed documents  
$$p(\beta_{1:K},\theta_{1:D},z_{1:D}|w_{1:D}) = \frac{p(\beta_{1:K},\theta_{1:D},z_{1:D},w_{1:D})}{p(w_{1:D})}.$$
The denominator of the posterior represents the probability of seeing the observed corpus under any topic model.  It is computed using the sampling based algorithm, Gibbs Sampling.

We generate topic models for the Hurricane Sandy and Katrina articles using LDA-C, developed by Blei in \cite{blei2003latent}.  We remove a list of common stop words from the corpus, along with common words specific to this corpus such as ``Sandy", ``Katrina", ``hurricane", and ``storm".  After filtering through the words, we use a Porter word stemmer to stem the remaining words, so each word is represented in one form, while it may appear in the articles in many different tenses \cite{porter1980algorithm}.

\subsection{Determining the Number of Topics}

The number of topics within a particular corpus depends on the size and scope of the corpus.  In our corpora, the scope is already quite narrow as we only focus on newspaper articles about a particular hurricane.  Thus, we do not expect the number of topics to be large, and to choose the number of topics for the analysis, we implement several techniques.

\begin{figure}[h]
\centerline{\includegraphics[width=0.5\textwidth]{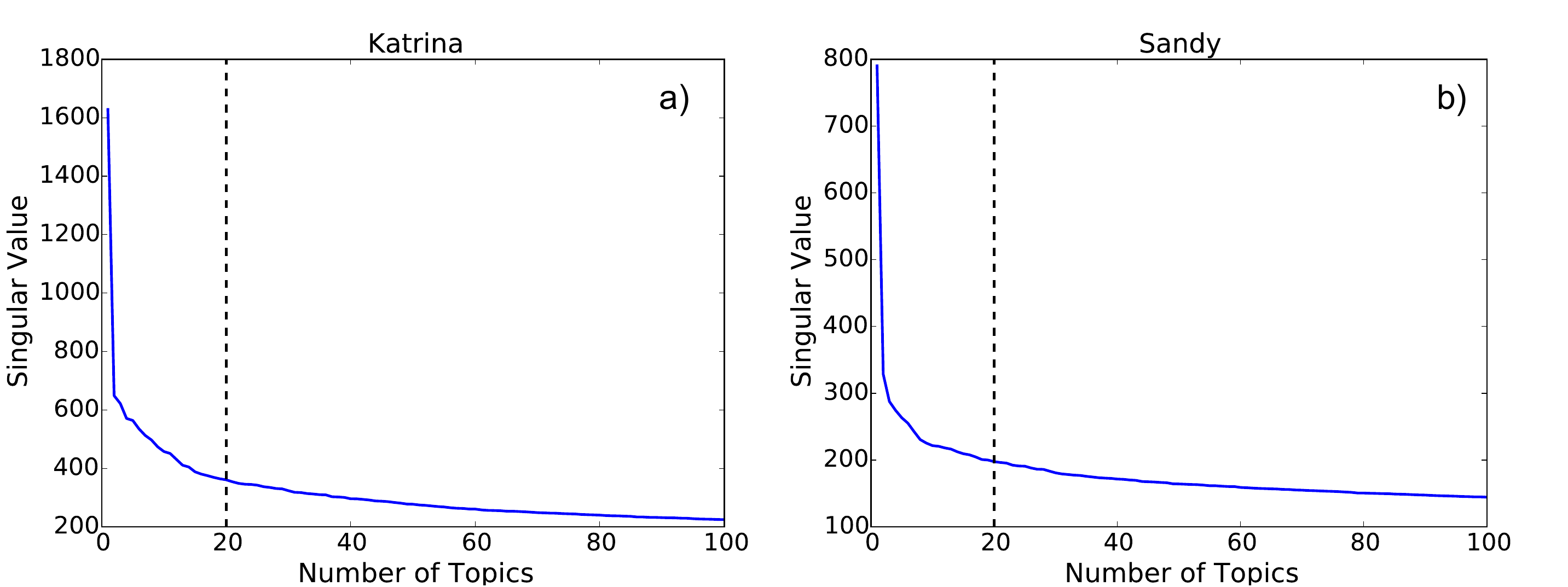}}
\caption{The 100 largest singular values in the (a) Hurricane Sandy and (b) Hurricane Katrina tf-idf matrices.  The elbow around 20 topics (see dashed line) determines the value of $k$ for SVD in LSA.}
\label{sing}
\end{figure}

First, to determine $k$, the rank of the approximated term-document matrix used in LSA, we look at the singular values determined via SVD.  The 100 largest singular values are plotted in Fig.~\ref{sing} for Hurricanes Sandy and Katrina.  The singular value decay rate slows considerably between singular values 20 and 30 for both matrices.  We find that topics become repetitive above $k=20$, and thus we choose $k=20$ as the rank of the approximated term-document matrix in LSA.

\begin{figure}[h] 
\centerline{\includegraphics[width=0.5\textwidth]{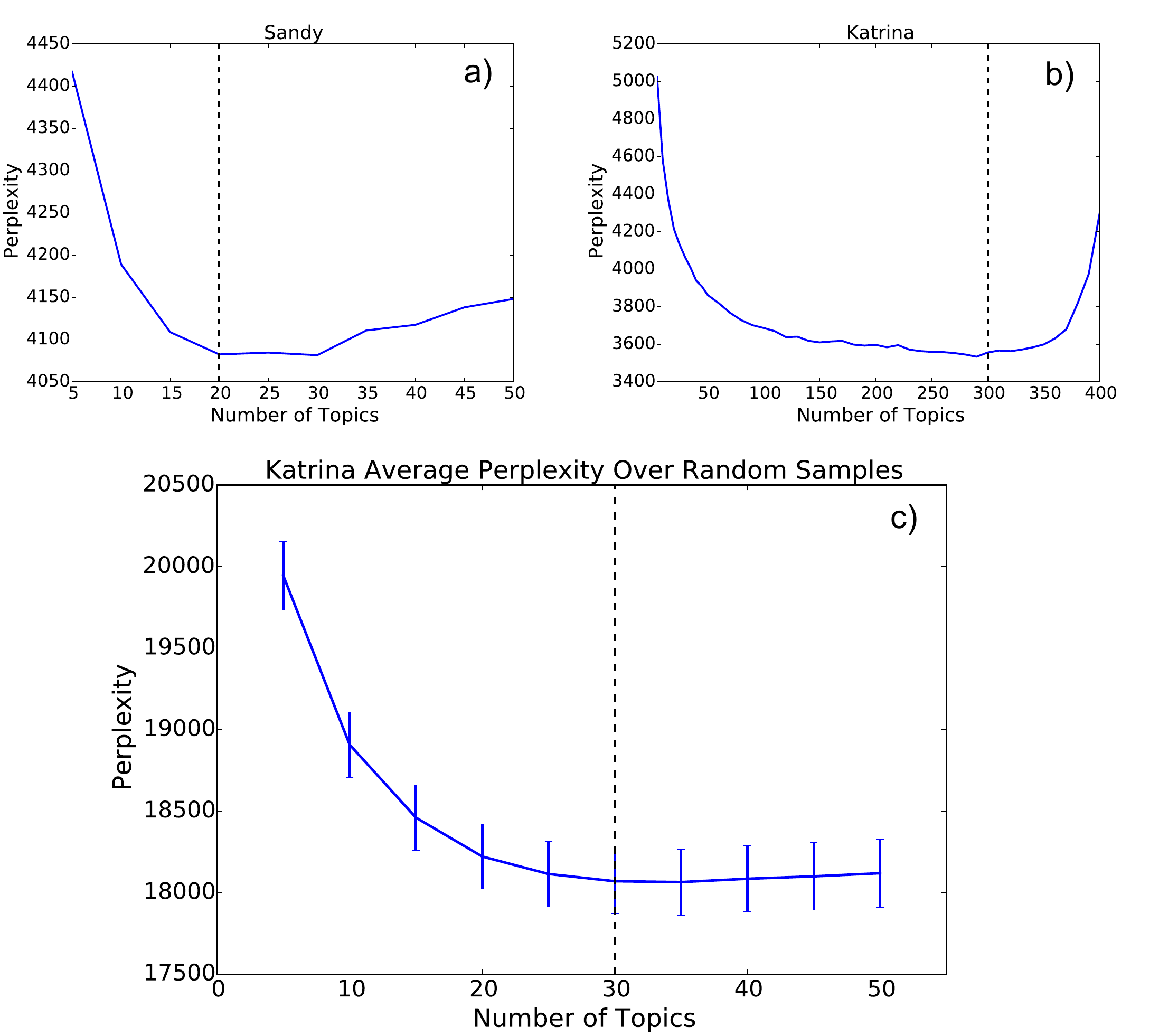}}
\caption{Average perplexity (over 10 testing sets) vs number of topics for the full (a) Sandy and (b) Katrina corpora.  Perplexity measures how well the model can predict a sample of unseen documents.  A lower perplexity indicates a better model.  Dashed lines show the optimal number of topics. (c) The average perplexity over 100 random samples of 1039 (the size of the Sandy corpus) documents from the Katrina corpus.  Each topic number is averaged first over 10 testing sets and then over 100 random samples from the full Katrina corpus.  Topic numbers increase by 2.  Error bars indicate the 95\% confidence intervals.}
\label{perplex}  
\end{figure}

To determine the number of topics for LDA to learn we use the perplexity, a measure employed in \cite{blei2003latent} to determine how accurately the topic model predicts a sample of unseen documents.  We compute the perplexity of a held out test set of documents for each hurricane, and vary the number of learned topics on the training data.  Perplexity will decrease with the number of topics and should eventually level out when increasing the number of topics no longer increases the accuracy of the model.  The perplexity may begin to increase when adding topics causes the model to overfit the data.  Perplexity is defined in \cite{blei2003latent} as
$$\textnormal{perplexity}(D_{\mathrm{test}}) = \mathrm{exp}\left\{-\frac{\sum_{d=1}^M\log p(\textbf{w}_d)}{\sum_{d=1}^MN_d}\right\},$$
where the numerator represents the log-likelihood of unseen documents $\textbf{w}_d$, and the denominator represents the total number of words in the testing set.  We separate the data into 10 equal testing and training sets for 10 fold cross validation on each hurricane.  We run LDA on each of the 10 different training sets consisting of 90\% of the articles in each hurricane corpus.  We then calculate the perplexity for a range of topic numbers on the testing sets, each consisting of 10\% of the articles.  We average the perplexity at each topic number over the testing sets, and plot the result in Fig.~\ref{perplex}(a) \& (b).

Figure \ref{perplex} indicates that the optimal number of topics in the Hurricane Sandy corpus is roughly 20 distinct topics, while the optimal number in the Hurricane Katrina corpus is between 280 and 300 distinct topics.  Compared to the Sandy corpus, the Hurricane Katrina corpus contains three times as many articles and about double the number of unique words (17,898 vs 9,521).  On average, an article in the Hurricane Sandy corpus contains 270 words, while an article in the Hurricane Katrina corpus contains 376 words.  The difference in these statistics may account for the difference in optimal topic numbers in Fig.~\ref{perplex}.  To test this hypothesis, we take 100 random samples of size 1039 (the size of the Sandy corpus) from the Katrina corpus and calculate the average perplexity over these samples.  For each of the 100 random samples, we use 10 testing and training sets for 10 fold cross validation, as was done in the previous calculations of perplexity.  We calculate the average perplexity over the 10 testing sets for each topic number, and then average over the 100 samples for each topic number, showing the result in Fig.~\ref{perplex}(c).  We find that on average, the optimal number of topics for a smaller Katrina corpus is around 30.     

Based on the above analysis, we opt to use a 20-topic model for Hurricane Sandy and a 30-topic model for Hurricane Katrina in our LDA analysis of the post-event media coverage.

\section{Results}
\label{results}
\subsection{Latent Semantic Analysis}
We compute a topic model for each corpus using LSA as described in the preceding methods section.  We provide 40 words most related to the three queries of interest in Tables \ref{LSAkatrina} \& \ref{LSAsandy}.  We list the 100 most related words to each query in the Supplementary Materials (see Tables \ref{LSAkatrinaFull} \& \ref{LSAsandyFull}).  While it is not possible to objectively explain why each word ranks where it does in the following lists, we search for a common theme within the words to determine how climate and energy were discussed in the media following these hurricanes.

\subsubsection{Hurricane Katrina}

\begin{table}
\centering
\resizebox{\columnwidth}{!}{%
\begin{tabular}[b]{cccccc}
\\\hline\noalign{\smallskip}
\multicolumn{6}{ c } {\textbf{Hurricane Katrina}} \\\hline\noalign{\smallskip}
``climate" & Similarity & ``energy" & Similarity& ``climate,energy" & Similarity  \\\noalign{\smallskip}\hline\noalign{\smallskip}
climate &	1.000 &	energy &	1.000 &	energy &	0.979 \\
larger &	0.866 &	prices &	0.986 &	prices &	0.952 \\
destroy &	0.861 &	exchange &	0.968 &	deutsche &	0.945 \\
formally &	0.848 &	consumers &	0.966 &	price &	0.943 \\
theory &	0.844 &	weinberg &	0.966 &	underinvestment &	0.943 \\
sound &	0.837 &	argus &	0.964 &	signaling &	0.941 \\
gale &	0.826 &	reidy &	0.962 &	discounting &	0.940 \\
reinforced &	0.817 &	splurge &	0.960 &	java &	0.940 \\
journal &	0.815 &	hummer &	0.960 &	argus &	0.939 \\
sensitive &	0.814 &	markets &	0.959 &	hummer &	0.938 \\
unlikely &	0.812 &	downers &	0.958 &	oil &	0.937 \\
belief &	0.809 &	highs &	0.958 &	consumers &	0.937 \\
phenomenon &	0.809 &	underinvestment &	0.957 &	shocks &	0.934 \\
rail &	0.800 &	exporting &	0.954 &	weinberg &	0.934 \\
studying &	0.796 &	price &	0.954 &	markets &	0.934 \\
wealthy &	0.795 &	reserves &	0.954 &	profits &	0.931 \\
brings &	0.792 &	signaling &	0.953 &	reserves &	0.931 \\
barge &	0.792 &	dampening &	0.950 &	exchange &	0.931 \\
ancient &	0.791 &	oil &	0.950 &	peaks &	0.931 \\
masters &	0.786 &	java &	0.949 &	highs &	0.929 \\
politicians &	0.785 &	cents &	0.948 &	splurge &	0.927 \\
professor &	0.783 &	deutsche &	0.948 &	exporting &	0.927 \\
recommendations &	0.782 &	gasoline &	0.947 &	gasoline &	0.923 \\
thick &	0.782 &	traders &	0.946 &	dampening &	0.923 \\
marked &	0.780 &	nariman &	0.946 &	pinch &	0.922 \\
alter &	0.779 &	discounting &	0.945 &	oils &	0.922 \\
sounds &	0.776 &	behravesh &	0.944 &	soaring &	0.922 \\
hole &	0.776 &	retailers &	0.943 &	exported &	0.920 \\
peril &	0.775 &	barrel &	0.942 &	reidy &	0.919 \\
extremely &	0.771 &	heating &	0.942 &	output &	0.919 \\
avoided &	0.770 &	oils &	0.942 &	exporter &	0.917 \\
loose &	0.770 &	shocks &	0.941 &	easing &	0.917 \\
multi &	0.769 &	idled &	0.941 &	putins &	0.917 \\
appear &	0.767 &	jolted &	0.941 &	record &	0.916 \\
devastating &	0.766 &	output &	0.940 &	tumbling &	0.916 \\
draft &	0.764 &	peaks &	0.937 &	demand &	0.915 \\
possibility &	0.764 &	profits &	0.936 &	downers &	0.915 \\
roiled &	0.759 &	soared &	0.936 &	automaker &	0.913 \\
retracted &	0.758 &	exported &	0.936 &	heating &	0.913 \\
mismanagement &	0.758 &	premcor &	0.935 &	disruptions &	0.913 \\\noalign{\smallskip}\hline
\end{tabular}
}
\caption{Results of LSA for Hurricane Katrina for 3 different queries.  Words are ordered based on their cosine similarity with the query vector.}
\label{LSAkatrina}
\end{table}

Within the Hurricane Katrina news media coverage, explicit reference to climate change was infrequent.  The set of words most related to ``climate" includes words such as ``theory", ``unlikely", ``belief", and ``possibility", indicating that linkages with climate change after Hurricane Katrina were tentative.  The uncertain link between hurricanes and climate change is often present in political discussions, thus the appearance of the word ``politician" in the ``climate" list is not surprising.  A direct quote from the article most related to the ``climate" query reads: 
\begin{quote}
``When two hurricanes as powerful as Katrina and Rita pummel the Gulf Coast so close together, many Americans are understandably wondering if something in the air has changed. Scientists are wondering the same thing. The field's leading researchers say it is too early to reach unequivocal conclusions. But some of them see evidence that global warming may be increasing the share of hurricanes that reach the monster magnitude of Katrina, and Rita'' \cite{lat}.
\end{quote} 
Words such as ``studying", ``professor", and ``masters" also indicate that reporting on climate change focused on research and academics.  The ``climate" list does not contain words relating to energy or energy systems and does not focus on the science or consequences of climate change. 

Within the 40 words most related to the ``energy" query, the majority pertain to energy prices and the stock market.  Within the ``climate" and ``energy" lists there is no overlap in the 40 most related words to these queries.

The ``climate" and ``energy" vectors are averaged to create the ``climate, energy" query vector.  The list of words most similar to this query is far more comparable to the ``energy" list than the ``climate" list.  Of the 100 most related words  to each query, there are 84 shared words between the ``energy" and ``climate, energy" lists.  This list again focuses on energy prices and not at all on climate change or infrastructure vulnerability, indicating that discussions about climate change, energy, and power outages were independent of one another within media reporting following Hurricane Katrina.  

\subsubsection{Hurricane Sandy}

\begin{table}
\centering
\resizebox{\columnwidth}{!}{%
\begin{tabular}[b]{cccccc}
\\\hline\noalign{\smallskip}
\multicolumn{6}{ c } {\textbf{Hurricane Sandy}} \\\hline\noalign{\smallskip}
``climate" & Similarity & ``energy" & Similarity & ``climate,energy" & Similarity  \\\noalign{\smallskip}\hline\noalign{\smallskip}
climate &	1.000 &	energy &	1.000 &	climate &	0.979 \\
change &	0.963 &	technologies &	0.949 &	warmer &	0.961 \\
reduce &	0.957 &	fuels &	0.946 &	georgetown &	0.956 \\
warming &	0.957 &	fossil &	0.943 &	warming &	0.955 \\
reducing &	0.956 &	hydroelectric &	0.936 &	reduce &	0.955 \\
pressures &	0.952 &	renewable &	0.932 &	energy &	0.952 \\
georgetown &	0.947 &	rogue &	0.932 &	reducing &	0.951 \\
lowering &	0.943 &	employing &	0.921 &	pressures &	0.948 \\
talks &	0.942 &	warmer &	0.920 &	fossil &	0.947 \\
devise &	0.938 &	supplying &	0.918 &	fuels &	0.946 \\
expands &	0.938 &	firing &	0.913 &	change &	0.946 \\
outweigh &	0.937 &	efficiency &	0.911 &	technologies &	0.945 \\
warmer &	0.937 &	streamlined &	0.911 &	coal &	0.943 \\
plants &	0.934 &	generating &	0.908 &	global &	0.942 \\
drought &	0.933 &	altering &	0.906 &	hydroelectric &	0.941 \\
manipulation &	0.929 &	coal &	0.906 &	emissions &	0.940 \\
emissions &	0.929 &	consumption &	0.900 &	firing &	0.937 \\
global &	0.929 &	adapt &	0.898 &	outweigh &	0.936 \\
imperative &	0.927 &	sparked &	0.895 &	generating &	0.933 \\
arizona &	0.924 &	dimming &	0.894 &	carbon &	0.930 \\
attribute &	0.923 &	georgetown &	0.892 &	arizona &	0.930 \\
scientists &	0.923 &	carbon &	0.889 &	editorials &	0.929 \\
planet &	0.920 &	masonry &	0.888 &	plants &	0.927 \\
pollution &	0.919 &	global &	0.886 &	humanitys &	0.926 \\
curbing &	0.918 &	erratic &	0.885 &	altering &	0.926 \\
coal &	0.917 &	searchable &	0.884 &	manipulation &	0.924 \\
editorials &	0.915 &	faster &	0.882 &	pollution &	0.923 \\
targets &	0.914 &	emissions &	0.881 &	employing &	0.923 \\
oceans &	0.912 &	skeptics &	0.880 &	drought &	0.922 \\
vigil &	0.912 &	proportion &	0.877 &	extracted &	0.921 \\
scenarios &	0.911 &	trillions &	0.876 &	foretaste &	0.920 \\
extracted &	0.911 &	foretaste &	0.876 &	skeptics &	0.919 \\
humanitys &	0.911 &	warming &	0.875 &	lowering &	0.919 \\
distraction &	0.910 &	reduce &	0.875 &	dioxide &	0.918 \\
pentagon &	0.910 &	editorials &	0.875 &	efficiency &	0.918 \\
contiguous &	0.909 &	humanitys &	0.875 &	planet &	0.917 \\
controlling &	0.908 &	eco &	0.875 &	curbing &	0.917 \\
carbon &	0.907 &	ton &	0.874 &	consumption &	0.915 \\
dioxide &	0.906 &	efficient &	0.872 &	expands &	0.914 \\
extremes &	0.905 &	cities &	0.872 &	subtler &	0.913 \\\noalign{\smallskip}\hline
\end{tabular}
}
\caption{Results of LSA for Hurricane Sandy for 3 different queries.  Words are ordered based on their cosine similarity with the query vector.}
\label{LSAsandy}
\end{table}

In the Hurricane Sandy corpus, we find the word ``climate" is most related to words describing climate change and global warming.  We also see words related to energy such as ``emissions", ``coal", ``carbon", and ``dioxide".  Including the top 100 words most related to ``climate" we see more energy related words including ``fossil", ``hydroelectric", ``technologies", and ``energy" itself.  This list differs substantially from that of the Hurricane Katrina analysis. 

The word ``energy" in the Hurricane Sandy corpus is most related to words describing climate change, such as the contributions of fossil fuels and the potential of renewable (``hydroelectric", ``renewable") energy resources.  This list of words focuses largely on how energy consumption is contributing to climate change, and, unlike the Katrina corpus, considerably overlaps with the list of ``climate" words.

Of the 100 words most related to ``energy", 58 of them are also listed in the 100 words most related to ``climate".  Of the 20 documents most related to the word ``energy", 15 of them are also listed in the 20 documents most related to ``climate".  Many of these articles discuss harmful emissions, renewable energy, and fossil fuels.     

In the Hurricane Sandy corpus, the ``climate, energy" query is again most related to the climate change and global warming related terms.  There are 87 shared terms in the ``climate" and ``climate, energy" lists and 66 shared terms in the ``energy" and ``climate, energy" related lists.  This result illustrates that when climate change was discussed in the media following Hurricane Sandy, energy related themes were often present.       

\subsection{Latent Dirichlet Allocation}
\begin{figure}[h!]
\centerline{\includegraphics[width=0.5\textwidth]{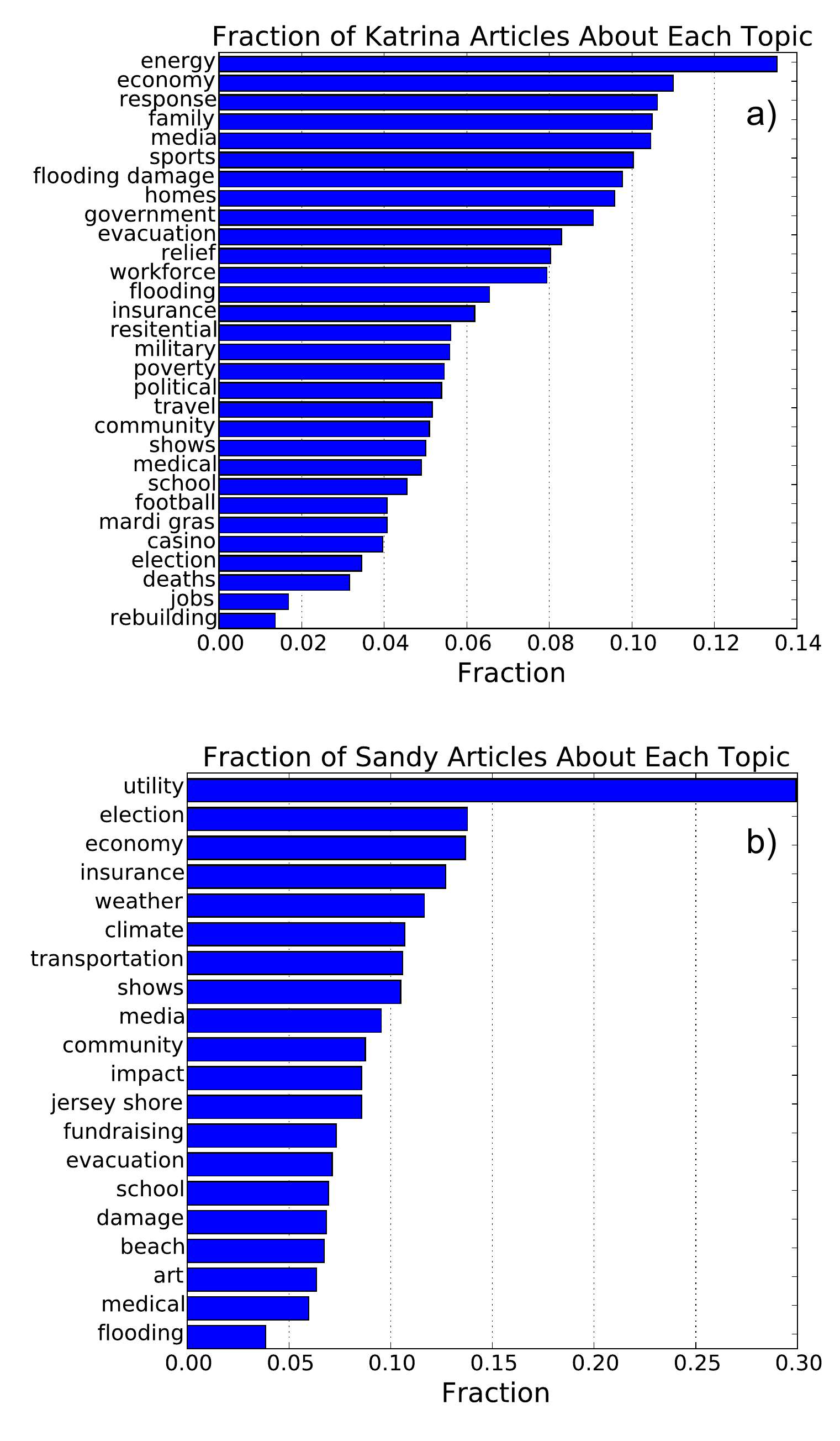}}
\caption{The proportion of articles ranking each topic as the first or second most probable topic, i.e., the proportion of articles that each topic appears in with high probability in the (a) Hurricane Katrina and (b) Hurricane Sandy corpora.  The topics order is by decreasing proportions.}
\label{doc_probs}  
\end{figure}

We generate LDA models for both the Sandy and Katrina corpora using 20 topics and 30 topics for Sandy and Katrina respectively (see Methods).  The 20 most probable words in 10 selected topic distributions are given in Tables \ref{LDAkatrina} \& \ref{LDAsandy}.  The full models are given in the Supplementary Materials (see Tables \ref{LDAkatrinaFull} \& \ref{LDAsandyFull}).  In addition to creating a distribution of topics over words,  LDA also creates a distribution of documents over topics.  Each topic is present in each document with some nonzero probability.  
We counted the number of times each topic appeared as one of the top two ranked topics in an article and divided this number by the number of articles in the corpus.  Fig.~\ref{doc_probs} summarizes the overall results of LDA for Katrina (a) and Sandy (b) by giving the proportion of articles that each topic appears in with high probability.  
We determined the topic names by manually analyzing the probability distribution of words in each topic.  We go into more detail on the topics of importance in the following sections. 

\subsubsection{Hurricane Katrina}

In Table \ref{LDAkatrina} we give 10 of the 30 topics in the LDA model for Hurricane Katrina.  In the Hurricane Katrina model, we see topics relating to deaths, relief, insurance, flooding, and energy.  We also see location specific topics such as sporting events, Mardi Gras, and music.  A major topic that is absent from this model is climate change.  Similar to the results we saw for the Katrina LSA model, the energy topic (Topic 8) in the Katrina LDA model contains words relating to energy prices, the market, and the economy.  In addition to a missing climate change topic, there is no mention of the climate within Topic 8 either, indicating that Hurricane Katrina did not only lack in climate change reporting but it also did not highlight the link between climate change and energy.   

\begin{table}
\centering
\resizebox{\columnwidth}{!}{%
\begin{tabular}[t]{ccccc}\\\hline\noalign{\smallskip}
\multicolumn{5}{ c } {\textbf{Hurricane Katrina}} \\\hline\noalign{\smallskip}
7: deaths &	8: energy &	12: relief &	13: family &	14: mardi gras \\\noalign{\smallskip}\hline\noalign{\smallskip}
   bodi &	   price &	   red &	   famili &	   gras \\
   death &	   oil &	   cross &	   home &	   mardi \\
   offici &	   percent &	   donat &	   children &	   french \\
   state &	   energi &	   relief &	   day &	   restaur \\
   home &	   gas &	   organ &	   live &	   parad \\
   die &	   gasolin &	   volunt &	   back &	   street \\
   victim &	   rate &	   victim &	   school &	   back \\
   peopl &	   market &	   fund &	   mother &	   peopl \\
   famili &	   week &	   peopl &	   friend &	   quarter \\
   parish &	   product &	   million &	   peopl &	   time \\
   st &	   month &	   chariti &	   im &	   home \\
   louisiana &	   consum &	   disast &	   call &	   day \\
   identifi &	   report &	   american &	   hous &	   citi \\
   morgu &	   economi &	   money &	   stay &	   make \\
   relat &	   compani &	   group &	   time &	   club \\
   coron &	   increas &	   rais &	   dont &	   louisiana \\
   dr &	   gulf &	   effort &	   work &	   cook \\
   dead &	   fuel &	   food &	   life &	   krew \\
   found &	   expect &	   org &	   son &	   hotel \\
   remain &	   gallon &	   shelter &	   left &	   celebr \\\hline\noalign{\smallskip}
16: shows &	19: travel &	21: insurance &	 28: evacuation & 29: response \\\noalign{\smallskip}\hline\noalign{\smallskip}
   music &	   ship &	   insur &	   hous &	   fema \\
   jazz &	   airlin &	   flood &	   evacue &	   respons \\
   band &	   show &	   damag &	   fema &	   feder \\
   musician &	   news &	   billion &	   peopl &	   agenc \\
   art &	   time &	   state &	   offici &	   brown \\
   cultur &	   northrop &	   compani &	   home &	   disast \\
   museum &	   network &	   loss &	   houston &	   govern \\
   perform &	   travel &	   mississippi &	   feder &	   emerg \\
   play &	   air &	   home &	   agenc &	   secur \\
   festiv &	   nbc &	   homeown &	   hotel &	   offici \\
   artist &	   million &	   pay &	   trailer &	   homeland \\
   song &	   broadcast &	   claim &	   famili &	   hous \\
   work &	   report &	   cost &	   state &	   depart \\
   show &	   abc &	   allstat &	   shelter &	   report \\
   time &	   cruis &	   area &	   emerg &	   manag \\
   concert &	   program &	   properti &	   live &	   chertoff \\
   includ &	   film &	   louisiana &	   month &	   white \\
   orchestra &	   channel &	   industri &	   apart &	   bush \\
   event &	   televis &	   feder &	   govern &	   plan \\
   record &	   navi &	   polici &	   assist &	   investig \\\noalign{\smallskip}\hline
\end{tabular}
}
\caption{The 20 most probable words within 10 of the 30 topic distributions given by LDA for Hurricane Katrina.  The words are stemmed according to a Porter stemmer \cite{porter1980algorithm}, where for example  \textit{flooded},  \textit{flooding}, and  \textit{floods} all become \textit{flood}.}
\label{LDAkatrina}
\end{table}

\subsubsection{Hurricane Sandy}
In the Hurricane Sandy LDA model, we see topics related to medics, insurance, fundraisers, government, damage, power outages, and climate change.  Unlike the Katrina model, we find that Topic 2 clearly represents climate change.  Words such as ``flood", ``weather", and ``natural" indicate that the reporting on climate change within articles about Hurricane Sandy discussed how climate change is contributing to weather extremes and natural disasters.  There was also considerable reporting on the rising sea levels, which are expected to contribute to the intensity of hurricanes and tropical storms \cite{michener1997climate}.

Dispersed throughout the weather related words in Topic 2, we see the words ``energy", ``power", and ``develop", indicating that power outages and energy system development were often discussed within articles that mentioned climate change, highlighting a link between climate change and the energy disruption caused by Hurricane Sandy.  Extending the number of words in Topic 2 we find more energy related words including ``infrastructure" (23), ``carbon" (28), ``resilience" (35), and ``emissions" (37).  A list of the 100 most probable words in Topic 2 is given in the Supplementary Information.  While ``carbon" and ``emissions" are clearly linked to climate change, words like ``infrastructure" and ``resilience" indicate a link between climate change discussion and energy system vulnerability. 

\begin{table}
\centering
\resizebox{\columnwidth}{!}{%
\begin{tabular}[t]{ccccc}\\\hline\noalign{\smallskip}
\multicolumn{5}{ c } {\textbf{Hurricane Sandy}} \\\hline\noalign{\smallskip}
\tb{0: utility} &	\tb{1: election} & \tb{2: climate} & \tb{3: community} &	\tb{7: transportation} \\\noalign{\smallskip}\hline\noalign{\smallskip}
   power &	   obama &	   climat &	   hous &	   train \\
   util &	   romney &	   flood &	   home &	   author \\
   servic &	   presid &	   chang &	   water &	   station \\
   compani &	   campaign &	   protect &	   beach &	   line \\
   author &	   elect &	   build &	   car &	   servic \\
   electr &	   state &	   rise &	   live &	   tunnel \\
   island &	   republican &	   sea &	   flood &	   jersey \\
   custom &	   vote &	   water &	   peopl &	   gas \\
   state &	   polit &	   risk &	   point &	   transport \\
   system &	   governor &	   level &	   fire &	   power \\
   grid &	   voter &	   energi &	   street &	   damag \\
   long &	   day &	   natur &	   rockaway &	   subway \\
   verizon &	   poll &	   power &	   back &	   street \\
   nation &	   democrat &	   weather &	   day &	   manhattan \\
   work &	   peopl &	   develop &	   insur &	   offici \\
   phone &	   debat &	   make &	   damag &	   transit \\
   commiss &	   candid &	   cost &	   resid &	   long \\
   network &	   presidenti &	   state &	   work &	   system \\
   con &	   time &	   plan &	   famili &	   day \\
   edison &	   nation &	   surg &	   neighborhood &	   island \\\hline\noalign{\smallskip}
\tb{8: medical} &	\tb{9: insurance} &	\tb{12: impact} &	\tb{13: media} &	\tb{15: fundraising }\\\noalign{\smallskip}\hline\noalign{\smallskip}
   hospit &	   insur &	   wind &	   show &	   concert \\
   home &	   compani &	   power &	   time &	   perform \\
   patient &	   percent &	   day &	   stewart &	   ticket \\
   health &	   sale &	   close &	   peopl &	   music \\
   medic &	   month &	   weather &	   make &	   show \\
   nurs &	   market &	   coast &	   photo &	   million \\
   evacu &	   busi &	   expect &	   live &	   money \\
   emerg &	   increas &	   servic &	   twitter &	   benefit \\
   center &	   million &	   travel &	   call &	   hall \\
   dr &	   loss &	   area &	   work &	   rais \\
   peopl &	   industri &	   offici &	   news &	   song \\
   citi &	   home &	   peopl &	   stori &	   peopl \\
   offici &	   report &	   state &	   includ &	   night \\
   resid &	   expect &	   damag &	   inform &	   work \\
   island &	   billion &	   flood &	   magazin &	   relief \\
   day &	   rate &	   nation &	   photograph &	   refund \\
   care &	   week &	   massachusett &	   design &	   springsteen \\
   bird &	   retail &	   center &	   post &	   jersey \\
   mayor &	   consum &	   report &	   print &	   sale \\
   mold &	   claim &	   hour &	   page &	   band \\\noalign{\smallskip}\hline
\end{tabular}
}
\caption{The 20 most probable words within 10 of the 20 topic distributions given by LDA for Hurricane Sandy.  The words are stemmed according to a Porter stemmer \cite{porter1980algorithm}.}
\label{LDAsandy}
\end{table}  

Topic 0 also contains words pertaining to energy systems.  This topic, however, does not contain any words pertaining to climate change.  Topic 0 is about electricity (``company", ``electricity", ``system"), power outages (``power",``utility", ``service"), and communication (``verizon", ``phone", ``network").  One benefit of LDA is that the model not only creates distributions of words over topics, but also distributions of topics over documents.  Of the 162 articles that are made up of more than 1\% Topic 2, 24 of them also contain Topic 0, demonstrating that these two topics were sporadically reported on in the same article.  For example, an article in \textit{The New York Times} entitled ``Experts Advise Cuomo on Disaster Measures" discusses how New York City can better prepare for drastic outages caused by extreme weather and directly quotes Governor Cuomo's concerns about climate change:
\begin{quote}
`` `Climate change is dramatically increasing the frequency and the severity of these situations,' Mr. Cuomo said. `And as time goes on, we're more and more realizing that these crises are more frequent and worse than anyone had predicted.' " \cite{nyt}
\end{quote}

Although the models for each hurricane generate some similar topics,  there are some topics in one model that do not appear in the other.  Both models give topics on politics, community, government aid, fundraisers, insurance, family, travel, medics, flooding, damage, evacuations, and energy.  The Hurricane Katrina model also gives topics relating to sporting events, Mardi Gras, music, military, and the death toll, while the Sandy model gives topics relating to museums, beaches, weather, Broadway, and climate change.  Many of the topics only appearing in one of the models appear there due to the hurricane's location.  The climate change topic, however, appears only in the Hurricane Sandy corpus and its absence in the Hurricane Katrina corpus cannot be simply be a consequence of the different locations of the hurricanes. 

\section{Discussion}
\label{discussion}

Through this analysis using topic models, we discover that climate change and energy were often discussed together within coverage of Hurricane Sandy, whereas the climate change topic is largely absent in post Hurricane Katrina reporting.  This difference can be attributed in part to changing public perceptions about climate change over time.  As early as 2001, the scientific consensus that climate change is occurring and resulting from human activity was legitimized by the IPCC assessment reports \cite{griggs2002climate}.  A 2003 national study on climate change risk perceptions, however, revealed that while most Americans demonstrate awareness of climate change, 68\% considered it only a moderate risk issue more likely to impact areas far from the United States \cite{leiserowitz2005american}.  In Fall 2008 (years after Hurricane Katrina), 51\% of Americans were either alarmed or concerned about global warming \cite{maibach2011identifying}, and in March 2012 (months before Hurricane Sandy), this number decreased to 39\% \cite{leiserowitz2012global}.  In April 2013, 38\% of Americans believed that people around the world are currently affected or harmed by the consequences of climate change \cite{leiserowitz2013climate}.  Those in the ``alarmed" and ``concerned" categories are also far more likely to report that they experienced a natural disaster within the last year \cite{leiserowitz2012global}, implying a potential relationship between personal experience of consequences and the perception of climate change risks \cite{myers2013relationship}.  Participants in the Yale School of Forestry \& Environmental Studies ``Americans and Climate Change" conference in 2005 determined that since science is the main source of climate change information, there is room for misinterpretation and disconnects in society's understanding of the issue \cite{yale}.

The 2004 and 2005 Atlantic hurricane seasons were among the costliest in United States history \cite{beven2008atlantic}.  In 2004, scientists began to propose that the intensity of the latest hurricane season may be linked to global warming.  However, the state of climate science at the time could not support such a hypothesis, and linkages between global warming and the impacts of hurricanes were deemed premature \cite{pielke2005hurricanes}.  Media coverage of climate change often presents  the scientific consensus and has influenced public opinion and risk perceptions on climate change \cite{antilla2008self}.  Complexity and uncertainty within the scientific community regarding the link between climate change and hurricanes may be why climate change does not appear as a prominent topic in the 2005 news media analysis of Hurricane Katrina.

Conversely, media reporting following Hurricane Sandy did connect explicitly with climate change.  By the time Hurricane Sandy occurred in 2012, climate science research had progressed and begun exploring the link between hurricanes and global warming \cite{mann2006atlantic, field2012managing, huber2011extreme}.  The Yale Project on Climate Change and Communications poll in March 2012 showed that a large majority of Americans believed at that time that certain weather extremes and natural disasters are caused by global warming \cite{leiserowitz2012extreme}.  This evolution of climate change research and public awareness is reflected in the different coverage of climate change after Hurricane Sandy.

Also unique to Hurricane Sandy coverage was the presence of climate and energy topics together.  While Hurricane Katrina reporting focused on the increase in energy prices following the storm, this increase in price was not explicitly linked to the consequences of climate change within media reporting.  Hurricane Katrina caused massive disruptions in oil and gas production in the Gulf of Mexico, which caused large spikes in the cost of oil and natural gas.  During Katrina, 2.6 million customers lost power in Louisiana, Mississippi, Alabama, Florida, and Georgia \cite{katrinaPower}.  The destruction caused by Katrina (followed shortly after by Hurricane Rita) encouraged drilling companies to upgrade their infrastructure to better withstand the forceful waves and wind from a large hurricane \cite{texas}.  During Hurricane Sandy, 8.66 million customers lost power from North Carolina to Maine, and it took 10 days for the utilities to restore power to 95\% of these affected customers.  Reporting on these outages is reflected in the LDA climate change topic.  Flooding and power outages at refineries, pipelines, and petroleum terminals in the New York Harbor area lead to gasoline shortages and prices increases \cite{sandyPower}.  These impacts illustrated some of the consequences of climate change and an increase in severity of natural disasters.  Hurricane Sandy news reporting not only highlighted the consequences of climate change but also the relationship between climate change, energy, and energy system vulnerability. 

\section{Conclusion}
\label{conclusion}

Given that the media both shapes and reflects public discourse, this analysis characterizing stark differences in media coverage between Hurricane Katrina and Hurricane Sandy demonstrates a shift in public discourse on climate change and energy systems. Although energy systems were disrupted in both storms, the connections between energy and climate change were made much more explicitly in the post-Hurricane Sandy news coverage as compared to the post-Hurricane Katrina coverage.  This shift is likely to represent multiple changes including: (1) increased public awareness and concern about climate change, (2) improved scientific understanding of the link between hurricane intensity and climate change, and (3) greater understanding of the energy system risks associated with climate change.  The ways that climate and energy are connected in the media coverage also reflects a larger shift toward increasing attention towards climate change adaptation in addition to climate mitigation \cite{hess2013transitions}.

Our investigation presents a mathematical approach to assessing public discourse of climate and energy, one that could be applied to assessing news media of other key areas in environmental studies. This analysis focuses on Hurricanes Katrina and Sandy due to their disruption and societal impact as focusing events.  Future  research  could expand to investigate how energy and climate are presented in other  climate and energy related media coverage over time.

\acknowledgments
We gratefully acknowledge support of this work from the National Science Foundation under project DGE-1144388, RAPID grant NSF-SES-1316442, and grant DMS-0940271 to the Mathematics \& Climate Research Network.  PSD was supported by NSF CAREER Grant No. 0846668.  Our work was crucially supported by the computational resources provided by the Vermont Advanced Computing Core and the Vermont Complex Systems Center.  We also acknowledge collaborators Tarla Rai Peterson, Elizabeth Wilson, Lauren Zeimer, and Andrea Feldpausch-Parker who contributed to initial post-Sandy media analysis.

\bibliographystyle{abbrv}


\begin{table*}
\resizebox{01.0\textwidth}{!}{
\centering
\begin{tabular}[c]{lccccccclcccccc}
\multicolumn{15}{ c } {\textbf{\LARGE{Supplementary Materials}}}\\\noalign{\smallskip}\noalign{\smallskip}\noalign{\smallskip}\noalign{\smallskip}\hline
\\\hline\noalign{\smallskip}
\multicolumn{15}{ c} {\textbf{Hurricane Katrina LSA}} \\\hline\noalign{\smallskip}
& ``climate" & Similarity & ``energy" &Similarity & ``climate, energy" &Similarity & & & ``climate" & Similarity & ``energy" &	Similarity & ``climate, energy" &Similarity\\\noalign{\smallskip}\hline\noalign{\smallskip}
1 &	climate &	1.000 &	energy &	1.000 &	energy &	0.979 &\hspace{3mm}	&	51 &	supposedly &	0.746 &	conocophillips &	0.929 &	retailers &	0.908 \\
2 &	larger &	0.866 &	prices &	0.986 &	prices &	0.952 &	&	52 &	boogie &	0.746 &	jumped &	0.928 &	citroen &	0.907 \\
3 &	destroy &	0.861 &	exchange &	0.968 &	deutsche &	0.945 &	&	53 &	theories &	0.746 &	citroen &	0.927 &	behravesh &	0.907 \\
4 &	formally &	0.848 &	consumers &	0.966 &	price &	0.943 &	&	54 &	nurtured &	0.745 &	tumbling &	0.926 &	traders &	0.906 \\
5 &	theory &	0.844 &	weinberg &	0.966 &	underinvestment &	0.943 &	&	55 &	raw &	0.745 &	mercantile &	0.925 &	producers &	0.905 \\
6 &	sound &	0.837 &	argus &	0.964 &	signaling &	0.941 &	&	56 &	topics &	0.744 &	production &	0.924 &	idled &	0.905 \\
7 &	gale &	0.826 &	reidy &	0.962 &	discounting &	0.940 &	&	57 &	sounded &	0.743 &	embargo &	0.923 &	products &	0.905 \\
8 &	reinforced &	0.817 &	splurge &	0.960 &	java &	0.940 &	&	58 &	cynthia &	0.742 &	putins &	0.922 &	tenth &	0.904 \\
9 &	journal &	0.815 &	hummer &	0.960 &	argus &	0.939 &	&	59 &	deadly &	0.742 &	shutdowns &	0.920 &	export &	0.904 \\
10 &	sensitive &	0.814 &	markets &	0.959 &	hummer &	0.938 &	&	60 &	sacrifice &	0.741 &	reserve &	0.920 &	commodity &	0.904 \\
11 &	unlikely &	0.812 &	downers &	0.958 &	oil &	0.937 &	&	61 &	certain &	0.740 &	crude &	0.920 &	imports &	0.903 \\
12 &	belief &	0.809 &	highs &	0.958 &	consumers &	0.937 &	&	62 &	cataclysmic &	0.740 &	arabica &	0.920 &	adjusting &	0.903 \\
13 &	phenomenon &	0.809 &	underinvestment &	0.957 &	shocks &	0.934 &	&	63 &	nor &	0.740 &	pretax &	0.919 &	yergin &	0.902 \\
14 &	rail &	0.800 &	exporting &	0.954 &	weinberg &	0.934 &	&	64 &	reconstructed &	0.739 &	mobil &	0.919 &	artificially &	0.902 \\
15 &	studying &	0.796 &	price &	0.954 &	markets &	0.934 &	&	65 &	assessments &	0.739 &	soaring &	0.919 &	nariman &	0.902 \\
16 &	wealthy &	0.795 &	reserves &	0.954 &	profits &	0.931 &	&	66 &	haunting &	0.738 &	uncharted &	0.919 &	cents &	0.902 \\
17 &	brings &	0.792 &	signaling &	0.953 &	reserves &	0.931 &	&	67 &	continuing &	0.737 &	imports &	0.919 &	tightness &	0.902 \\
18 &	barge &	0.792 &	dampening &	0.950 &	exchange &	0.931 &	&	68 &	transforming &	0.737 &	chevrons &	0.919 &	subjective &	0.902 \\
19 &	ancient &	0.791 &	oil &	0.950 &	peaks &	0.931 &	&	69 &	william &	0.737 &	exxon &	0.917 &	doha &	0.901 \\
20 &	masters &	0.786 &	java &	0.949 &	highs &	0.929 &	&	70 &	regard &	0.736 &	manifold &	0.917 &	spikes &	0.900 \\
21 &	politicians &	0.785 &	cents &	0.948 &	splurge &	0.927 &	&	71 &	vicinity &	0.736 &	trading &	0.916 &	winter &	0.898 \\
22 &	professor &	0.783 &	deutsche &	0.948 &	exporting &	0.927 &	&	72 &	booming &	0.735 &	suisse &	0.916 &	exxon &	0.897 \\
23 &	recommendations &	0.782 &	gasoline &	0.947 &	gasoline &	0.923 &	&	73 &	audiences &	0.735 &	automaker &	0.916 &	uncharted &	0.897 \\
24 &	thick &	0.782 &	traders &	0.946 &	dampening &	0.923 &	&	74 &	advocacy &	0.734 &	tepid &	0.915 &	chairmans &	0.897 \\
25 &	marked &	0.780 &	nariman &	0.946 &	pinch &	0.922 &	&	75 &	mass &	0.733 &	futures &	0.915 &	soared &	0.897 \\
26 &	alter &	0.779 &	discounting &	0.945 &	oils &	0.922 &	&	76 &	remarkable &	0.733 &	geopolitical &	0.915 &	conocophillips &	0.896 \\
27 &	sounds &	0.776 &	behravesh &	0.944 &	soaring &	0.922 &	&	77 &	breaking &	0.732 &	record &	0.914 &	clamping &	0.895 \\
28 &	hole &	0.776 &	retailers &	0.943 &	exported &	0.920 &	&	78 &	facts &	0.732 &	yergin &	0.914 &	exporters &	0.895 \\
29 &	peril &	0.775 &	barrel &	0.942 &	reidy &	0.919 &	&	79 &	constituents &	0.731 &	clamping &	0.914 &	bps &	0.895 \\
30 &	extremely &	0.771 &	heating &	0.942 &	output &	0.919 &	&	80 &	isolated &	0.730 &	retail &	0.914 &	crimp &	0.895 \\
31 &	avoided &	0.770 &	oils &	0.942 &	exporter &	0.917 &	&	81 &	vibrant &	0.703 &	hess &	0.913 &	cutback &	0.894 \\
32 &	loose &	0.770 &	shocks &	0.941 &	easing &	0.917 &	&	82 &	unequivocal &	0.730 &	pinch &	0.912 &	global &	0.894 \\
33 &	multi &	0.769 &	idled &	0.941 &	putins &	0.917 &	&	83 &	recommended &	0.729 &	chairmans &	0.911 &	pretax &	0.893 \\
34 &	appear &	0.767 &	jolted &	0.941 &	record &	0.916 &	&	84 &	unprotected &	0.728 &	closings &	0.911 &	disrupted &	0.893 \\
35 &	devastating &	0.766 &	output &	0.940 &	tumbling &	0.916 &	&	85 &	inundated &	0.727 &	depository &	0.910 &	liquefied &	0.892 \\
36 &	draft &	0.764 &	peaks &	0.937 &	demand &	0.915 &	&	86 &	ears &	0.726 &	disrupted &	0.909 &	premcor &	0.892 \\
37 &	possibility &	0.764 &	profits &	0.936 &	downers &	0.915 &	&	87 &	exuberant &	0.725 &	winter &	0.909 &	jumped &	0.891 \\
38 &	roiled &	0.759 &	soared &	0.936 &	automaker &	0.913 &	&	88 &	greenhouse &	0.725 &	sunoco &	0.909 &	mobil &	0.891 \\
39 &	retracted &	0.758 &	exported &	0.936 &	heating &	0.913 &	&	89 &	powers &	0.725 &	chevron &	0.908 &	arabica &	0.890 \\
40 &	mismanagement &	0.758 &	premcor &	0.935 &	disruptions &	0.913 &	&	90 &	alarms &	0.724 &	doha &	0.908 &	bros &	0.890 \\
41 &	plot &	0.757 &	disruptions &	0.934 &	atm &	0.911 &	&	91 &	comment &	0.723 &	bros &	0.907 &	mercantile &	0.890 \\
42 &	produced &	0.757 &	exporter &	0.934 &	tepid &	0.911 &	&	92 &	brokers &	0.722 &	commodity &	0.907 &	analyst &	0.887 \\
43 &	becomes &	0.755 &	easing &	0.933 &	chevrons &	0.911 &	&	93 &	deny &	0.722 &	commodities &	0.906 &	gas &	0.887 \\
44 &	decades &	0.753 &	crimp &	0.932 &	jolted &	0.911 &	&	94 &	pianos &	0.722 &	wholesalers &	0.905 &	geopolitical &	0.886 \\
45 &	consider &	0.752 &	dent &	0.932 &	embargo &	0.909 &	&	95 &	baker &	0.721 &	refiner &	0.905 &	interruptions &	0.886 \\
46 &	wealthier &	0.752 &	demand &	0.932 &	pricing &	0.909 &	&	96 &	ethnic &	0.720 &	soar &	0.903 &	squeeze &	0.886 \\
47 &	dismissed &	0.751 &	roasters &	0.930 &	roasters &	0.908 &	&	97 &	cyclical &	0.720 &	analyst &	0.903 &	chevron &	0.885 \\
48 &	repeated &	0.751 &	tightness &	0.930 &	dent &	0.908 &	&	98 &	relieve &	0.720 &	products &	0.903 &	crude &	0.885 \\
49 &	delays &	0.750 &	atm &	0.929 &	production &	0.908 &	&	99 &	studies &	0.720 &	bps &	0.903 &	nations &	0.884 \\
50 &	unique &	0.749 &	pricing &	0.929 &	barrel &	0.908 &	&	100 &	spread &	0.720 &	thurtell &	0.902 &	derivatives &	0.884 \\\noalign{\smallskip}\hline
\end{tabular}}
\caption{Results of LSA for Hurricane Katrina for 3 different queries.  Words are ordered based on their cosine distance from the query vector.  Includes the 100 words most similar to the query.}
\label{LSAkatrinaFull}
\end{table*}

\begin{table*}
\resizebox{1.0\textwidth}{!}{
\centering
\begin{tabular}[c]{lccccccclcccccc}
\\\hline\noalign{\smallskip}
\multicolumn{15}{ c } {\textbf{Hurricane Sandy LSA}} \\\hline\noalign{\smallskip}
& ``climate" & Similarity & ``energy" &Similarity & ``climate, energy" &Similarity & & & ``climate" & Similarity & ``energy" &	Similarity & ``climate, energy" &Similarity\\\noalign{\smallskip}\hline\noalign{\smallskip}
1 &	climate &	1.000 &	energy &	1.000 &	climate &	0.979 &\hspace{3mm}	&	51 &	fuels &	0.895 &	extracted &	0.862 &	deniers &	0.906 \\
2 &	change &	0.963 &	technologies &	0.949 &	warmer &	0.961 &	&	52 &	kerry &	0.894 &	abundance &	0.860 &	vigil &	0.904 \\
3 &	reduce &	0.957 &	fuels &	0.946 &	georgetown &	0.956 &	&	53 &	hydroelectric &	0.893 &	tackle &	0.860 &	proportion &	0.904 \\
4 &	warming &	0.957 &	fossil &	0.943 &	warming &	0.955 &	&	54 &	pollute &	0.893 &	regulating &	0.858 &	targets &	0.902 \\
5 &	reducing &	0.956 &	hydroelectric &	0.936 &	reduce &	0.955 &	&	55 &	technologies &	0.891 &	outweigh &	0.858 &	mover &	0.901 \\
6 &	pressures &	0.952 &	renewable &	0.932 &	energy &	0.952 &	&	56 &	altering &	0.890 &	envisioned &	0.857 &	scientists &	0.899 \\
7 &	georgetown &	0.947 &	rogue &	0.932 &	reducing &	0.951 &	&	57 &	regulating &	0.890 &	miserably &	0.856 &	automobiles &	0.899 \\
8 &	lowering &	0.943 &	employing &	0.921 &	pressures &	0.948 &	&	58 &	mover &	0.889 &	subtler &	0.856 &	devise &	0.898 \\
9 &	talks &	0.942 &	warmer &	0.920 &	fossil &	0.947 &	&	59 &	believing &	0.886 &	upending &	0.855 &	controlling &	0.898 \\
10 &	devise &	0.938 &	supplying &	0.918 &	fuels &	0.946 &	&	60 &	enhancement &	0.885 &	pollution &	0.855 &	modification &	0.896 \\
11 &	expands &	0.938 &	firing &	0.913 &	change &	0.946 &	&	61 &	planets &	0.885 &	solar &	0.855 &	trillions &	0.895 \\
12 &	outweigh &	0.937 &	efficiency &	0.911 &	technologies &	0.945 &	&	62 &	eco &	0.883 &	modification &	0.855 &	scenarios &	0.893 \\
13 &	warmer &	0.937 &	streamlined &	0.911 &	coal &	0.943 &	&	63 &	cities &	0.882 &	sciences &	0.854 &	earths &	0.893 \\
14 &	plants &	0.934 &	generating &	0.908 &	global &	0.942 &	&	64 &	automobiles &	0.882 &	automobiles &	0.853 &	abundance &	0.891 \\
15 &	drought &	0.933 &	altering &	0.906 &	hydroelectric &	0.941 &	&	65 &	greenhouse &	0.880 &	regulatory &	0.852 &	attribute &	0.890 \\
16 &	manipulation &	0.929 &	coal &	0.906 &	emissions &	0.940 &	&	66 &	notoriously &	0.879 &	trapping &	0.851 &	greenhouse &	0.888 \\
17 &	emissions &	0.929 &	consumption &	0.900 &	firing &	0.937 &	&	67 &	strict &	0.878 &	surprises &	0.850 &	enhancement &	0.887 \\
18 &	global &	0.929 &	adapt &	0.898 &	outweigh &	0.936 &	&	68 &	porous &	0.878 &	earths &	0.849 &	doom &	0.886 \\
19 &	imperative &	0.927 &	sparked &	0.895 &	generating &	0.933 &	&	69 &	groundwater &	0.878 &	measured &	0.848 &	funneling &	0.885 \\
20 &	arizona &	0.924 &	dimming &	0.894 &	carbon &	0.930 &	&	70 &	consumption &	0.877 &	mover &	0.847 &	groundwater &	0.885 \\
21 &	attribute &	0.923 &	georgetown &	0.892 &	arizona &	0.930 &	&	71 &	modification &	0.876 &	waterkeeper &	0.846 &	hotter &	0.884 \\
22 &	scientists &	0.923 &	carbon &	0.889 &	editorials &	0.929 &	&	72 &	hotter &	0.876 &	change &	0.845 &	copenhagen &	0.884 \\
23 &	planet &	0.920 &	masonry &	0.888 &	plants &	0.927 &	&	73 &	earths &	0.875 &	deniers &	0.843 &	oceans &	0.883 \\
24 &	pollution &	0.919 &	global &	0.886 &	humanitys &	0.926 &	&	74 &	markedly &	0.875 &	sub &	0.843 &	windstorms &	0.883 \\
25 &	curbing &	0.918 &	erratic &	0.885 &	altering &	0.926 &	&	75 &	retaining &	0.875 &	blackouts &	0.843 &	planets &	0.881 \\
26 &	coal &	0.917 &	searchable &	0.884 &	manipulation &	0.924 &	&	76 &	attests &	0.875 &	manipulation &	0.842 &	emission &	0.881 \\
27 &	editorials &	0.915 &	faster &	0.882 &	pollution &	0.923 &	&	77 &	dimming &	0.875 &	depleting &	0.841 &	munich &	0.881 \\
28 &	targets &	0.914 &	emissions &	0.881 &	employing &	0.923 &	&	78 &	employing &	0.874 &	funneling &	0.841 &	rogue &	0.880 \\
29 &	oceans &	0.912 &	skeptics &	0.880 &	drought &	0.922 &	&	79 &	proportion &	0.873 &	curbing &	0.841 &	markedly &	0.878 \\
30 &	vigil &	0.912 &	proportion &	0.877 &	extracted &	0.921 &	&	80 &	efficiency &	0.873 &	plants &	0.841 &	pollute &	0.878 \\
31 &	scenarios &	0.911 &	trillions &	0.876 &	foretaste &	0.920 &	&	81 &	depleted &	0.873 &	sources &	0.841 &	ozone &	0.878 \\
32 &	extracted &	0.911 &	foretaste &	0.876 &	skeptics &	0.919 &	&	82 &	exemplified &	0.872 &	frequent &	0.841 &	depleting &	0.877 \\
33 &	humanitys &	0.911 &	warming &	0.875 &	lowering &	0.919 &	&	83 &	murky &	0.872 &	oil &	0.841 &	epa &	0.877 \\
34 &	distraction &	0.910 &	reduce &	0.875 &	dioxide &	0.918 &	&	84 &	sparked &	0.870 &	planet &	0.839 &	overheated &	0.877 \\
35 &	pentagon &	0.910 &	editorials &	0.875 &	efficiency &	0.918 &	&	85 &	essay &	0.870 &	emission &	0.838 &	contiguous &	0.877 \\
36 &	contiguous &	0.909 &	humanitys &	0.875 &	planet &	0.917 &	&	86 &	atmospheric &	0.869 &	ozone &	0.837 &	frequent &	0.876 \\
37 &	controlling &	0.908 &	eco &	0.875 &	curbing &	0.917 &	&	87 &	overheated &	0.869 &	pentagon &	0.836 &	sensible &	0.874 \\
38 &	carbon &	0.907 &	ton &	0.874 &	consumption &	0.915 &	&	88 &	copenhagen &	0.869 &	windstorms &	0.836 &	freely &	0.872 \\
39 &	dioxide &	0.906 &	efficient &	0.872 &	expands &	0.914 &	&	89 &	fahrenheit &	0.868 &	acceptance &	0.836 &	kerry &	0.871 \\
40 &	extremes &	0.905 &	cities &	0.872 &	subtler &	0.913 &	&	90 &	energy &	0.868 &	buildup &	0.835 &	ton &	0.870 \\
41 &	munich &	0.903 &	doom &	0.870 &	dimming &	0.912 &	&	91 &	adapt &	0.868 &	copenhagen &	0.835 &	fahrenheit &	0.870 \\
42 &	firing &	0.902 &	compounding &	0.869 &	talks &	0.911 &	&	92 &	windstorms &	0.867 &	focuses &	0.835 &	exemplified &	0.869 \\
43 &	subtler &	0.902 &	mentioning &	0.868 &	sparked &	0.910 &	&	93 &	funneling &	0.867 &	vein &	0.834 &	persistence &	0.869 \\
44 &	foretaste &	0.900 &	climate &	0.868 &	pentagon &	0.909 &	&	94 &	illustrative &	0.866 &	epa &	0.834 &	atmospheric &	0.869 \\
45 &	generating &	0.899 &	reducing &	0.867 &	eco &	0.909 &	&	95 &	vapor &	0.866 &	drought &	0.833 &	environmental &	0.866 \\
46 &	environmental &	0.899 &	pressures &	0.866 &	adapt &	0.909 &	&	96 &	abundance &	0.864 &	harvard &	0.832 &	increasing &	0.865 \\
47 &	fossil &	0.899 &	arizona &	0.864 &	imperative &	0.908 &	&	97 &	prosperity &	0.864 &	redundant &	0.832 &	levi &	0.865 \\
48 &	deniers &	0.898 &	candlelit &	0.863 &	trapping &	0.908 &	&	98 &	freely &	0.863 &	greenhouse &	0.829 &	meaningfully &	0.864 \\
49 &	trapping &	0.897 &	dioxide &	0.862 &	cities &	0.907 &	&	99 &	emission &	0.862 &	temperature &	0.827 &	porous &	0.863 \\
50 &	skeptics &	0.896 &	degrees &	0.862 &	regulating &	0.906 &	&	100 &	scientific &	0.862 &	iron &	0.826 &	essay &	0.862 \\\noalign{\smallskip}\hline
\end{tabular}
}
\caption{Results of LSA for Hurricane Sandy for 3 different queries.  Words are ordered based on their cosine distance from the query vector.  Includes the 100 words most similar to the query.}
\label{LSAsandyFull}
\end{table*}

\begin{table*}
\resizebox{0.8\textwidth}{!}{
\centering
\begin{tabular}[t]{cccccccccc}\\\hline\noalign{\smallskip}
\multicolumn{10}{ c } {\textbf{Hurricane Katrina LDA}} \\\hline\noalign{\smallskip}
topic 0 &	topic 1 &	topic 2 &	topic 3 &	topic 4 &	topic 5 &	topic 6 &	topic 7 &	topic 8 &	topic 9 \\\noalign{\smallskip}\hline\noalign{\smallskip}
   quinn &	   leve &	   job &	   hous &	   billion &	   polic &	   bush &	   bodi &	   price &	   school \\
   team &	   corp &	   hous &	   water &	   tax &	   casino &	   presid &	   death &	   oil &	   student \\
   season &	   engin &	   krt &	   home &	   feder &	   offic &	   democrat &	   offici &	   percent &	   univers \\
   time &	   flood &	   st &	   street &	   hous &	   peopl &	   republican &	   state &	   energi &	   tulan \\
   player &	   water &	   home &	   time &	   senat &	   street &	   hous &	   home &	   gas &	   colleg \\
   play &	   canal &	   antoin &	   day &	   congress &	   day &	   polit &	   die &	   gasolin &	   educ \\
   game &	   protect &	   back &	   peopl &	   cut &	   depart &	   white &	   victim &	   rate &	   back \\
   coach &	   wall &	   school &	   back &	   republican &	   fire &	   administr &	   peopl &	   market &	   campus \\
   start &	   system &	   restaur &	   tree &	   spend &	   citi &	   senat &	   famili &	   week &	   return \\
   point &	   louisiana &	   peopl &	   live &	   bill &	   biloxi &	   respons &	   parish &	   product &	   high \\
   open &	   armi &	   work &	   boat &	   budget &	   hotel &	   govern &	   st &	   month &	   district \\
   make &	   offici &	   month &	   resid &	   govern &	   crime &	   nation &	   louisiana &	   consum &	   enrol \\
   made &	   surg &	   worker &	   work &	   money &	   store &	   american &	   identifi &	   report &	   public \\
   day &	   feet &	   louisiana &	   build &	   program &	   reddick &	   time &	   morgu &	   economi &	   class \\
   sign &	   project &	   day &	   neighborhood &	   state &	   water &	   leader &	   relat &	   compani &	   warm \\
   top &	   level &	   live &	   damag &	   propos &	   time &	   critic &	   coron &	   increas &	   research \\
   week &	   pump &	   chitrib &	   roof &	   cost &	   back &	   peopl &	   dr &	   gulf &	   time \\
   score &	   lake &	   rebuild &	   photograph &	   bush &	   gambl &	   iraq &	   dead &	   fuel &	   hurrican \\
   world &	   design &	   end &	   flood &	   plan &	   mississippi &	   parti &	   found &	   expect &	   teacher \\
   lead &	   environment &	   return &	   photo &	   million &	   hous &	   effort &	   remain &	   gallon &	   institut \\\hline\noalign{\smallskip}
topic 10 &	topic 11 &	topic 12 &	topic 13 &	topic 14 &	topic 15 &	topic 16 &	topic 17 &	topic 18 &	topic 19 \\\noalign{\smallskip}\hline\noalign{\smallskip}
   peopl &	   leve &	   red &	   famili &	   gras &	   game &	   music &	   town &	   peopl &	   ship \\
   black &	   hous &	   cross &	   home &	   mardi &	   team &	   jazz &	   plan &	   time &	   airlin \\
   king &	   flood &	   donat &	   children &	   french &	   saint &	   band &	   build &	   american &	   show \\
   time &	   protect &	   relief &	   day &	   restaur &	   play &	   musician &	   develop &	   disast &	   news \\
   west &	   rebuild &	   organ &	   live &	   parad &	   season &	   art &	   school &	   news &	   time \\
   mayor &	   home &	   volunt &	   back &	   street &	   home &	   cultur &	   hous &	   report &	   northrop \\
   day &	   system &	   victim &	   school &	   back &	   footbal &	   museum &	   state &	   world &	   network \\
   presid &	   feder &	   fund &	   mother &	   peopl &	   player &	   perform &	   design &	   stori &	   travel \\
   polit &	   work &	   peopl &	   friend &	   quarter &	   coach &	   play &	   communiti &	   book &	   air \\
   bloomberg &	   offici &	   million &	   peopl &	   time &	   state &	   festiv &	   resid &	   nation &	   nbc \\
   democrat &	   peopl &	   chariti &	   im &	   home &	   time &	   artist &	   architect &	   thing &	   million \\
   campaign &	   hotel &	   disast &	   call &	   day &	   leagu &	   song &	   board &	   public &	   broadcast \\
   franklin &	   state &	   american &	   hous &	   citi &	   stadium &	   work &	   meet &	   word &	   report \\
   candid &	   neighborhood &	   money &	   stay &	   make &	   giant &	   show &	   public &	   natur &	   abc \\
   ferrer &	   engin &	   group &	   time &	   club &	   san &	   time &	   local &	   day &	   cruis \\
   poll &	   corp &	   rais &	   dont &	   louisiana &	   back &	   concert &	   street &	   govern &	   program \\
   hop &	   powel &	   effort &	   work &	   cook &	   bowl &	   includ &	   architectur &	   media &	   film \\
   made &	   billion &	   food &	   life &	   krew &	   louisiana &	   orchestra &	   project &	   great &	   channel \\
   hip &	   busi &	   org &	   son &	   hotel &	   field &	   event &	   urban &	   make &	   televis \\
   dont &	   krt &	   shelter &	   left &	   celebr &	   win &	   record &	   peopl &	   histori &	   navi \\\hline\noalign{\smallskip}
topic 20 &	topic 21 &	topic 22 &	topic 23 &	topic 24 &	topic 25 &	topic 26 &	topic 27 &	topic 28 &	topic 29 \\\noalign{\smallskip}\hline\noalign{\smallskip}
   compani &	   insur &	   peopl &	   hospit &	   guard &	   nagin &	   evacu &	   state &	   hous &	   fema \\
   busi &	   flood &	   church &	   patient &	   nation &	   neighborhood &	   water &	   car &	   evacue &	   respons \\
   work &	   damag &	   black &	   health &	   state &	   resid &	   peopl &	   charg &	   fema &	   feder \\
   employe &	   billion &	   massachusett &	   medic &	   militari &	   black &	   offici &	   law &	   peopl &	   agenc \\
   million &	   state &	   state &	   nurs &	   troop &	   mayor &	   resid &	   court &	   offici &	   brown \\
   contract &	   compani &	   poverti &	   care &	   offici &	   citi &	   louisiana &	   vehicl &	   home &	   disast \\
   servic &	   loss &	   work &	   dr &	   bush &	   rebuild &	   rita &	   investig &	   houston &	   govern \\
   worker &	   mississippi &	   romney &	   center &	   unit &	   white &	   area &	   offic &	   feder &	   emerg \\
   bank &	   home &	   poor &	   doctor &	   forc &	   peopl &	   flood &	   attorney &	   agenc &	   secur \\
   custom &	   homeown &	   american &	   peopl &	   feder &	   elect &	   coast &	   lawyer &	   hotel &	   offici \\
   week &	   pay &	   evacue &	   evacu &	   louisiana &	   home &	   state &	   louisiana &	   trailer &	   homeland \\
   oper &	   claim &	   servic &	   state &	   equip &	   vote &	   texa &	   case &	   famili &	   hous \\
   port &	   cost &	   communiti &	   flu &	   day &	   hous &	   center &	   judg &	   state &	   depart \\
   system &	   allstat &	   job &	   emerg &	   effort &	   area &	   wind &	   report &	   shelter &	   report \\
   execut &	   area &	   base &	   staff &	   relief &	   flood &	   emerg &	   fraud &	   emerg &	   manag \\
   line &	   properti &	   day &	   home &	   presid &	   return &	   gulf &	   station &	   live &	   chertoff \\
   area &	   louisiana &	   time &	   day &	   respons &	   plan &	   home &	   feder &	   month &	   white \\
   damag &	   industri &	   live &	   die &	   blanco &	   percent &	   day &	   offici &	   apart &	   bush \\
   small &	   feder &	   worker &	   diseas &	   disast &	   lower &	   houston &	   file &	   govern &	   plan \\
   call &	   polici &	   nation &	   univers &	   rescu &	   landrieu &	   mile &	   system &	   assist &	   investig \\\noalign{\smallskip}\hline
\end{tabular}
}
\caption{A 30 topic LDA model for Hurricane Katrina.  Each topic contains the 20 most probable (stemmed) words in its distribution.  We stem words according to a Porter stemmer \cite{porter1980algorithm}.}
\label{LDAkatrinaFull}
\end{table*}

\begin{table*}
\resizebox{1.0\textwidth}{!}{
\centering
\begin{tabular}[t]{cccccccccc}\\\hline\noalign{\smallskip}
\multicolumn{10}{ c } {\textbf{Hurricane Sandy LDA}} \\\hline\noalign{\smallskip}
topic 0 &	topic 1 &	topic 2 &	topic 3 &	topic 4 &	topic 5 &	topic 6 &	topic 7 &	topic 8 &	topic 9 \\\noalign{\smallskip}\hline\noalign{\smallskip}
   power &	   obama &	   climat &	   hous &	   school &	   broadway &	   park &	   train &	   hospit &	   insur \\
   util &	   romney &	   flood &	   home &	   time &	   street &	   tree &	   author &	   home &	   compani \\
   servic &	   presid &	   chang &	   water &	   fund &	   theater &	   boardwalk &	   station &	   patient &	   percent \\
   compani &	   campaign &	   protect &	   beach &	   peopl &	   time &	   jersey &	   line &	   health &	   sale \\
   author &	   elect &	   build &	   car &	   day &	   work &	   damag &	   servic &	   medic &	   month \\
   electr &	   state &	   rise &	   live &	   student &	   open &	   fire &	   tunnel &	   nurs &	   market \\
   island &	   republican &	   sea &	   flood &	   children &	   perform &	   seasid &	   jersey &	   evacu &	   busi \\
   custom &	   vote &	   water &	   peopl &	   public &	   peopl &	   shore &	   gas &	   emerg &	   increas \\
   state &	   polit &	   risk &	   point &	   famili &	   day &	   busi &	   transport &	   center &	   million \\
   system &	   governor &	   level &	   fire &	   american &	   show &	   height &	   power &	   dr &	   loss \\
   grid &	   voter &	   energi &	   street &	   red &	   week &	   summer &	   damag &	   peopl &	   industri \\
   long &	   day &	   natur &	   rockaway &	   donat &	   power &	   town &	   subway &	   citi &	   home \\
   verizon &	   poll &	   power &	   back &	   case &	   run &	   time &	   street &	   offici &	   report \\
   nation &	   democrat &	   weather &	   day &	   work &	   danc &	   beach &	   manhattan &	   resid &	   expect \\
   work &	   peopl &	   develop &	   insur &	   cross &	   play &	   work &	   offici &	   island &	   billion \\
   phone &	   debat &	   make &	   damag &	   govern &	   light &	   pier &	   transit &	   day &	   rate \\
   commiss &	   candid &	   cost &	   resid &	   live &	   night &	   island &	   long &	   care &	   week \\
   network &	   presidenti &	   state &	   work &	   disast &	   cancel &	   stand &	   system &	   bird &	   retail \\
   con &	   time &	   plan &	   famili &	   parent &	   halloween &	   back &	   day &	   mayor &	   consum \\
   edison &	   nation &	   surg &	   neighborhood &	   relief &	   close &	   visit &	   island &	   mold &	   claim \\\hline\noalign{\smallskip}
topic 10 &	topic 11 &	topic 12 &	topic 13 &	topic 14 &	topic 15 &	topic 16 &	topic 17 &	topic 18 &	topic 19 \\\noalign{\smallskip}\hline\noalign{\smallskip}
   museum &	   hous &	   wind &	   show &	   peopl &	   concert &	   feder &	   water &	   beach &	   build \\
   art &	   water &	   power &	   time &	   home &	   perform &	   billion &	   system &	   sand &	   street \\
   work &	   peopl &	   day &	   stewart &	   live &	   ticket &	   state &	   state &	   island &	   develop \\
   galleri &	   build &	   close &	   peopl &	   hous &	   music &	   hous &	   million &	   park &	   apart \\
   water &	   resid &	   weather &	   make &	   water &	   show &	   aid &	   flood &	   dune &	   properti \\
   street &	   home &	   coast &	   photo &	   hotel &	   million &	   disast &	   plant &	   long &	   million \\
   damag &	   volunt &	   expect &	   live &	   day &	   money &	   money &	   cost &	   offici &	   floor \\
   flood &	   food &	   servic &	   twitter &	   polic &	   benefit &	   program &	   car &	   rockaway &	   estat \\
   space &	   day &	   travel &	   call &	   work &	   hall &	   damag &	   occupi &	   corp &	   water \\
   center &	   work &	   area &	   work &	   resid &	   rais &	   govern &	   sewag &	   project &	   manhattan \\
   compani &	   power &	   offici &	   news &	   famili &	   song &	   republican &	   river &	   debri &	   flood \\
   build &	   live &	   peopl &	   stori &	   apart &	   peopl &	   jersey &	   peopl &	   town &	   resid \\
   seaport &	   red &	   state &	   includ &	   time &	   night &	   million &	   dutch &	   home &	   real \\
   includ &	   island &	   damag &	   inform &	   island &	   work &	   congress &	   project &	   resid &	   owner \\
   insur &	   apart &	   flood &	   magazin &	   door &	   relief &	   cuomo &	   build &	   communiti &	   squar \\
   offic &	   week &	   nation &	   photograph &	   evacu &	   refund &	   senat &	   geotherm &	   sea &	   damag \\
   artist &	   street &	   massachusett &	   design &	   call &	   springsteen &	   insur &	   work &	   day &	   tenant \\
   aquarium &	   heat &	   center &	   post &	   worker &	   jersey &	   cost &	   park &	   boardwalk &	   month \\
   site &	   brooklyn &	   report &	   print &	   staten &	   sale &	   offici &	   area &	   public &	   feet \\
   research &	   hook &	   hour &	   page &	   report &	   band &	   homeown &	   engin &	   work &	   move \\\noalign{\smallskip}\hline
\end{tabular}
}
\caption{A 20 topic LDA model for Hurricane Sandy.  Each topic contains the 20 most probable words in its distribution.  We stem words according to a Porter stemmer \cite{porter1980algorithm}.}
\label{LDAsandyFull}
\end{table*}

\begin{table*}
\resizebox{.8\textwidth}{!}{
\centering
\begin{tabular}[t]{cccccccccccccc}\\\hline\noalign{\smallskip}
\multicolumn{4}{ c } {\textbf{Sandy Topic 0}} & & \multicolumn{4}{ c } {\textbf{Sandy Topic 2}} & & \multicolumn{4}{ c } {\textbf{Katrina Topic 8}}\\\noalign{\smallskip}\hline\noalign{\smallskip}   
1 &	power &	51 &	   generat &	&	1 &	   climat &	51 &	   coastal &	&	1 &	   price &	51 &	   drop \\
2 &	   util &	52 &	   solar &	&	2 &	   flood &	52 &	   bloomberg &	&	2 &	   oil &	52 &	   reserv \\
3 &	   servic &	53 &	   spokesman &	&	3 &	   chang &	53 &	   public &	&	3 &	   percent &	53 &	   close \\
4 &	   compani &	54 &	   voic &	&	4 &	   protect &	54 &	   barrier &	&	4 &	   energi &	54 &	   inflat \\
5 &	   author &	55 &	   energi &	&	5 &	   build &	55 &	   part &	&	5 &	   gas &	55 &	   spend \\
6 &	   electr &	56 &	   manag &	&	6 &	   rise &	56 &	   elev &	&	6 &	   gasolin &	56 &	   refin \\
7 &	   island &	57 &	   emerg &	&	7 &	   sea &	57 &	   presid &	&	7 &	   rate &	57 &	   august \\
8 &	   custom &	58 &	   liberti &	&	8 &	   water &	58 &	   system &	&	8 &	   market &	58 &	   depart \\
9 &	   state &	59 &	   local &	&	9 &	   risk &	59 &	   map &	&	9 &	   week &	59 &	   natur \\
10 &	   system &	60 &	   respons &	&	10 &	   level &	60 &	   gas &	&	10 &	   product &	60 &	   chief \\
11 &	   grid &	61 &	   governor &	&	11 &	   energi &	61 &	   vulner &	&	11 &	   month &	61 &	   job \\
12 &	   long &	62 &	   prepar &	&	12 &	   natur &	62 &	   disast &	&	12 &	   consum &	62 &	   end \\
13 &	   verizon &	63 &	   feder &	&	13 &	   power &	63 &	   peopl &	&	13 &	   report &	63 &	   septemb \\
14 &	   nation &	64 &	   copper &	&	14 &	   weather &	64 &	   fuel &	&	14 &	   economi &	64 &	   profit \\
15 &	   work &	65 &	   govern &	&	15 &	   develop &	65 &	   event &	&	15 &	   compani &	65 &	   feder \\
16 &	   phone &	66 &	   rate &	&	16 &	   make &	66 &	   polici &	&	16 &	   increas &	66 &	   gain \\
17 &	   commiss &	67 &	   problem &	&	17 &	   cost &	67 &	   step &	&	17 &	   gulf &	67 &	   retail \\
18 &	   network &	68 &	   regul &	&	18 &	   state &	68 &	   zone &	&	18 &	   fuel &	68 &	   record \\
19 &	   con &	69 &	   link &	&	19 &	   plan &	69 &	   damag &	&	19 &	   expect &	69 &	   interest \\
20 &	   edison &	70 &	   cost &	&	20 &	   surg &	70 &	   live &	&	20 &	   gallon &	70 &	   damag \\
21 &	   day &	71 &	   percent &	&	21 &	   nation &	71 &	   effect &	&	21 &	   cent &	71 &	   share \\
22 &	   restor &	72 &	   backup &	&	22 &	   warm &	72 &	   unit &	&	22 &	   barrel &	72 &	   rais \\
23 &	   public &	73 &	   report &	&	23 &	   infrastructur &	73 &	   coast &	&	23 &	   higher &	73 &	   term \\
24 &	   communic &	74 &	   investig &	&	24 &	   global &	74 &	   research &	&	24 &	   stock &	74 &	   demand \\
25 &	   million &	75 &	   damag &	&	25 &	   increas &	75 &	   agenc &	&	25 &	   economist &	75 &	   futur \\
26 &	   worker &	76 &	   chief &	&	26 &	   citi &	76 &	   recent &	&	26 &	   quarter &	76 &	   billion \\
27 &	   execut &	77 &	   elli &	&	27 &	   reduc &	77 &	   long &	&	27 &	   econom &	77 &	   level \\
28 &	   cuomo &	78 &	   ed &	&	28 &	   carbon &	78 &	   generat &	&	28 &	   high &	78 &	   declin \\
29 &	   offici &	79 &	   wireless &	&	29 &	   environment &	79 &	   heat &	&	29 &	   cost &	79 &	   hit \\
30 &	   employe &	80 &	   carrier &	&	30 &	   scientist &	80 &	   effort &	&	30 &	   suppli &	80 &	   investor \\
31 &	   batteri &	81 &	   presid &	&	31 &	   billion &	81 &	   rais &	&	31 &	   day &	81 &	   survey \\
32 &	   offic &	82 &	   hit &	&	32 &	   engin &	82 &	   pollut &	&	32 &	   analyst &	82 &	   state \\
33 &	   oper &	83 &	   counti &	&	33 &	   studi &	83 &	   industri &	&	33 &	   refineri &	83 &	   remain \\
34 &	   week &	84 &	   general &	&	34 &	   time &	84 &	   project &	&	34 &	   nation &	84 &	   effect \\
35 &	   call &	85 &	   consum &	&	35 &	   resili &	85 &	   standard &	&	35 &	   industri &	85 &	   hurrican \\
36 &	   time &	86 &	   consolid &	&	36 &	   futur &	86 &	   code &	&	36 &	   time &	86 &	   impact \\
37 &	   charg &	87 &	   equip &	&	37 &	   emiss &	87 &	   hit &	&	37 &	   rose &	87 &	   heat \\
38 &	   provid &	88 &	   director &	&	38 &	   area &	88 &	   issu &	&	38 &	   point &	88 &	   credit \\
39 &	   plan &	89 &	   issu &	&	39 &	   govern &	89 &	   ocean &	&	39 &	   rise &	89 &	   labor \\
40 &	   includ &	90 &	   cabl &	&	40 &	   feet &	90 &	   oyster &	&	40 &	   fell &	90 &	   servic \\
41 &	   pay &	91 &	   critic &	&	41 &	   requir &	91 &	   design &	&	41 &	   fed &	91 &	   american \\
42 &	   wire &	92 &	   cellphon &	&	42 &	   higher &	92 &	   larg &	&	42 &	   averag &	92 &	   continu \\
43 &	   statu &	93 &	   technolog &	&	43 &	   mayor &	93 &	   offici &	&	43 &	   trade &	93 &	   unit \\
44 &	   failur &	94 &	   run &	&	44 &	   extrem &	94 &	   warn &	&	44 &	   million &	94 &	   show \\
45 &	   home &	95 &	   caus &	&	45 &	   propos &	95 &	   face &	&	45 &	   growth &	95 &	   produc \\
46 &	   line &	96 &	   telephon &	&	46 &	   high &	96 &	   east &	&	46 &	   index &	96 &	   note \\
47 &	   board &	97 &	   substat &	&	47 &	   plant &	97 &	   sever &	&	47 &	   yesterday &	97 &	   petroleum \\
48 &	   panel &	98 &	   guard &	&	48 &	   includ &	98 &	   univers &	&	48 &	   sale &	98 &	   earn \\
49 &	   hour &	99 &	   place &	&	49 &	   insur &	99 &	   decad &	&	49 &	   crude &	99 &	   concern \\
50 &	   area &	100 &	   chairman &	&	50 &	   world &	100 &	   solut &	&	50 &	   coast &	100 &	   import \\\noalign{\smallskip}\hline
\end{tabular}
}
\caption{A 100 word extension of selected topics from the Sandy and Katrina LDA models.}
\label{LSAextended}
\end{table*}
\end{document}